\documentclass[aps,twocolumn,prl,showpacs,superscriptaddress,notitlepage,longbibliography]{revtex4-1}

\usepackage{graphicx}
\usepackage{epsfig}
\usepackage{amsmath}

\usepackage{graphics}
\usepackage{latexsym}
\usepackage{color}
\usepackage{amssymb}
\usepackage{bm}
\usepackage{yhmath}

\usepackage{url,enumitem}

\begin{document}

\title{Number-parity effect for confined fermions in one dimension}
\author{Christian Schilling}
\email{christian.schilling@physics.ox.ac.uk}
\affiliation{Clarendon Laboratory, University of Oxford, Parks Road, Oxford OX1 3PU, United Kingdom}
\author{Rolf Schilling}
\email{rschill@uni-mainz.de}
\affiliation{Institut f\"ur Physik, Johannes Gutenberg-Universit\"at Mainz, Staudinger Weg 9, D-55099 Mainz, Germany}

\begin{abstract}
For $N$ spin-polarized fermions with harmonic pair interactions in a $1$-dimensional trap an odd-even effect is found. The spectrum of the $1$-particle reduced density matrix of the system's ground state differs qualitatively  for $N$ odd and $N$ even. This effect does only occur for strong attractive and repulsive interactions. Since it does not exists for bosons, it must originate from the repulsive nature implied by the fermionic exchange statistics.
In contrast to the spectrum, the $1$-particle density and correlation function  for strong attractive interactions do not show any sensitivity on the number parity. This also suggests that reduced-density-matrix-functional theory has a more subtle $N$-dependency than density functional theory.
\end{abstract}
\maketitle

\paragraph{Introduction.---}
Physical behavior often depends qualitatively on binary parameters as, e.g.,~ odd/even or
integer/half-integer. Such parity effects play an important role in physics.
A well-known example is Kramers' number-parity effect, i.e.~the twofold degeneracy of the eigenstates of a quantum system with an odd number of electrons, provided time reversal symmetry holds \cite{Kramers}. Haldane \cite{Haldaneparity1} has shown the existence of a spin-parity effect. The spectrum of the quantum Heisenberg antiferromagnet in one dimension has an energy gap for all integer spins whereas it is gapless in case of half integer spins.
Recently, an interesting number-parity effect has been observed experimentally for a few ultracold fermions in a quasi-one-dimensional trap. Tuning the potential such that the pair interactions become attractive, Cooper pairs are formed. Their tunneling is different for an odd and an even number of fermions \cite{Jochimpairing}. Based on Kramers' theorem another number-parity effect was proven to exist  for fermionic $1$-particle reduced density matrices ($1$-RDM) \cite{Smith1966} (see also Ref.~\cite{Col2}). There it was shown that the eigenvalues of a $1$-RDM arising from an eigenstate of a time reversal symmetric hamiltonian are twofold degenerate for an even number of fermions.

In the present work we will show that the  so-called natural occupation numbers, i.e., the spectrum  $\{\lambda_{k}\}$ of the ground state $1$-RDM for strongly attractive and spinpolarized fermions confined in one dimension also exhibits an odd-even-effect. Since the spin-polarizing magnetic field breaks time reversal symmetry, this effect is completely different from that found in Ref.~\cite{Smith1966}. Furthermore, it does not occur for bosons. Consequently, it must result from the fermionic exchange symmetry.

Besides the relevance of parity effects on their own, exploring the structure of reduced density matrices has also gained a lot of relevance during
recent years. This is essentially due to progress \cite{Kly4,Daft,Kly2,MC2006,Kly3,Altun} in the quantum marginal problem (QMP) which studies the relation of reduced density matrices arising from a common multipartite quantum state. For basic overviews of the QMP the reader may consult Refs.~\cite{CSQMath12,CSthesis,Alex}. The most prominent QMP is the $2$-particle $N$-representability problem \cite{Col2,Dav}, the description of $2$-RDM arising from $N$-fermion quantum states. Its solution would allow to efficiently calculate ground states of fermionic quantum systems with $2$-body interactions.
However, since this problem and most of the other QMP are Quantum-Merlin-Arthur-hard \cite{QMA}, already partial insights on the set of compatible density matrices are highly appreciated and alternative methods for the ground state calculation are gaining importance as well. One such promising method is reduced-density-matrix-functional theory (see e.g.~\cite{Gilbert,Col3}). This natural extension of density functional theory \cite{Hohenberg} seeks a distinguished functional $\mathcal{F}$ on the $1$-RDM whose minimization leads to the exact ground state energy and the corresponding ground state $1$-RDM.
Any structural insights on ground state $1$-RDM  contributes to this task of finding or approximating $\mathcal{F}$ by exposing further necessary constraints on legitimate functionals.

\paragraph{Model and $1$-RDM.---}
We consider $N$ identical particles with mass $m$ in a one-dimensional harmonic trap interacting via a harmonic two-body potential. If $x_i$  is the position of the $i$-th particle the hamiltonian reads:
\begin{equation} \label{eq1}
\hat{H}=\sum\limits_{i=1}^N \Big[- \frac{\hbar^2}{2m} \frac{\partial^2}{\partial x^2_i} + \frac{1}{2} m \omega^2 x^2_i\Big]+ \frac{1}{2} D\! \sum\limits_{1 \leq i < j \leq N} \Big(x_i-x_j \Big)^2\,,
\end{equation}
where $\omega$ is the eigenfrequency of the trap and $D$ the interaction strength, which can be positive or negative. Stability requires $D > D_{low}\equiv - m \omega^2/N$.

Hamiltonian (\ref{eq1}) arises as an effective model, e.g.,~for the description of quantum dots, where the Coulomb interaction between the electrons is screened (see e.g.~Ref.~\cite{QuantumDot}). Furthermore, it was used to understand the emergence of shell structures in atoms (see e.g.~Ref.~\cite{HarmShells}) and nuclei (see e.g.~Ref.~\cite{ZinnerNucl}). 

The great advantage of model (\ref{eq1}) is the exact knowledge of all its eigenstates \cite{harmOsc2012,CS2013,CS2013NO,BECNON,ZinnerNbody}.
For arbitrary numbers of bosons and any spatial dimension the ground state $1$-RDM can easily be calculated \cite{CS2013,CS2013NO,BECNON,ZinnerNbody}.
Based on such analytic results Bose-Einstein condensation in harmonic traps was explored (see, e.g.,~Ref.~\cite{HarmBEC}). In contrast to bosons, the analytical calculation of the corresponding fermionic $1$-RDM is much more involved. We will show that the properties of the fermionic 1-RDM are much richer, compared to the bosonic case, leading to new insights.

For \textit{spin-polarized} (or \textit{spinless}) fermions in one dimension, the spatial part of the ground state of hamiltonian (\ref{eq1}) is the totally antisymmetric wave function \cite{harmOsc2012,CS2013,CS2013NO} (see also Refs.~\cite{HarmShells,QuantumDot})
\begin{eqnarray} \label{eq2}
\Psi_0^{(f)} (x_1, \ldots, x_N) = \mathcal{N}_{N} ^{(f)}\cdot \Big[\prod_{1 \leq i < j \leq N} (x_i - x_j)\Big] \hspace{0.5cm} &&\nonumber \\
\hspace{0.0cm}\cdot \exp [-A(x^2_1 + \ldots + x^2_N) + B (x_1 + \ldots + x_N)^2 ] &&
\end{eqnarray}
with $\mathcal{N}_{N}^{(f)}$ a normalization factor and
\begin{equation} \label{eq3}
A=\frac{1}{2 l^2_+} \, , \quad B=\frac{1}{2N} \Big(\frac{1}{l^2_+} - \frac{1}{l^2_-}\Big) \,.
\end{equation}
$l_-=\sqrt{\hbar/m \omega}$ and $l_+=\sqrt{\hbar/m(\omega^2 + DN/m)^{1/2}}$ are the length scales for the center of mass and the relative motion, respectively. Note that $\Psi_0^{(f)} (x_1, \ldots, x_N)$ resembles Laughlin's wave function \cite{Laughlin1983} for the fractional quantum Hall effect. We will come back to this point below.

The $1$-RDM of (\ref{eq2}) follows as \cite{CS2013,CS2013NO}
\begin{equation} \label{eq4}
\rho^{(f)}_{1}(x;y) =\mathcal{N}^{(f)} F(x,y) \exp [-a (x^2 + y^2) + b xy]
\end{equation}
with the polynomial $F(x,y)$ in $x$ and $y$ of degree $2(N-1)$ (see Appendix \ref{app:FandV}) and
\begin{equation} \label{eq8}
a=A-B-b/2 \quad , \quad b=(N-1) B^2/[A-(N-1)B]  \,.
\end{equation}
The factor $\mathcal{N}^{(f)}$ (not to be confused with $\mathcal{N}_{N}^{(f)}$, the normalization constant of $\Psi_0^{(f)}$),
follows from the normalization $\int\limits^\infty_{- \infty} dx \rho_1^{(f)} (x;x) = N$. The bosonic and fermionic $1$-RDM are related by \cite{CS2013NO}
\begin{equation} \label{eq9}
\rho^{(f)}_1 (x;y)= \mathcal{ \tilde{N}}^{(f)} F(x,y)\, \rho^{(b)}_1 (x;y) \,,
\end{equation}
where $\mathcal{ \tilde{N}}^{(f)}=\mathcal{N}^{(f)}/\mathcal{N}^{(b)}$ and $\mathcal{N}^{(b)}$
the corresponding normalization constant appearing in $\rho^{(b)}_1 (x;y)$.
Consequently, the fermionic nature of $\rho^{(f)}_1 (x;y)$ is only contained in the polynomial prefactor $F(x,y)$. It arises from the polynomial prefactor of the exponential function in Eq.~(\ref{eq2}), the Vandermonde determinant, which is a result of the fermionic exchange symmetry.

The  spectrum $\{\lambda_k\}$  follows from solving the eigenvalue equation

\begin{equation} \label{eq10}
\int\limits_{- \infty}^\infty dy \rho^{(f)}_1 (x;y) \chi^{(f)}_k (y)= \lambda^{(f)}_k \chi^{(f)}_k (x) \,,
\end{equation}
$k=1,2,3,\ldots$. In quantum chemistry the $\lambda_k$ are called natural occupation numbers.  These eigenvalues will be  ordered decreasingly, i.e.,~$\lambda_k \geq \lambda_{k+1}$ for all $k \geq 1$.
Due to the duality $\{\lambda_k(l_-,l_+)\}=\{\lambda_k(l_+,l_-)\}$, first observed in Ref.~\cite{Nagydual1} and proven in Ref.~\cite{duality}, we restrict to $\l_+/l_- \leq 1$. Notice, $\l_+/l_- < 1$ ($\l_+/l_- > 1$) means attractive (repulsive) pair interactions.
Since we are interested only in fermions we also suppress the superscript $(f)$.

\paragraph{Strong coupling limit and results.---}

For \textit{attractive} interaction the strong coupling limit $DN/m \omega^2 \rightarrow \infty$ corresponds to the limit $t \equiv l_+/l_- \rightarrow 0$, performed at $N$ fixed. For \textit{repulsive} interactions  it follows for $DN/m \omega^2 \rightarrow D_{low}N/m \omega^2$, corresponding to $t \rightarrow \infty$.

We will prove that completely unexpected features of the  spectrum of the fermionic $1$-RDM occur in this limit, which can be discussed analytically. Due to the duality property \cite{duality} we can restrict to attractive interactions, i.e. to $t \leq 1$. Technical details can be found in the appendix. 

In the following all coordinates $x_{i}$ and all lengths will be measured in units of $l_-$.
For  $t \rightarrow 0 $, there are three basic observations.  First, the leading order of $\rho_1(x;y)$  is proportional to
$\exp \Big[-\frac{N-1}{4N}((x-y)/t)^2 \Big]$ (see Appendix \ref{app:NONs}). Therefore, the weight of the deviation of $y$ from $x$ decreases extremely fast for decreasing $t$. Second, $\chi_{k}(x)$ varies on an $x$-scale proportional to $\sqrt{t}$ and third, the polynomial prefactor $F(x,y)$ converges to a polynomial $\tilde{F}(\tilde{z})$ where $\tilde{z}=(x-y)/t$ (ses Appendix \ref{app:1rdm}).
Let us introduce the rescaled variable  $\tilde{x}= x/\sqrt{t}$ and $\chi_{k}(\sqrt{t}\tilde{x})= \exp(\frac{1}{2}Nt \tilde{x}^2) \zeta_{k}(\tilde{x})$.  Then, by using the momentum representation the eigenvalue equation (\ref{eq10}) for $t \ll 1$ reduces to a Schr\"odinger equation (with position and momentum exchanged)
\begin{equation} \label{eq11}
\Big[ - \frac{\hbar^2}{2 \tilde{m}} \frac{\partial^2}{\partial \tilde{p}^2} + V_{N}(\tilde{p})\Big] \tilde{\zeta_{k}}(\tilde{p}) = (- \lambda_{k}) \ \tilde{\zeta_{k}}(\tilde{p}) \ .
\end{equation}
Here $\tilde{\zeta_{k}}(\tilde{p})$ is the Fourier transform of $\zeta_{k}(\tilde{x})$,
\begin{equation} \label{eq12}
\tilde{m}_N= \hbar^2/(-2t N V_{N}(0))
\end{equation}
the mass of the `particle' and
\begin{equation} \label{eq13}
V_N(\tilde{p})=- t  \mathcal{N}  \int\limits_{- \infty}^\infty d \tilde{z} \tilde{F}(\tilde{z}) \cos(\sqrt{t}\tilde{p}\tilde{z})\exp\Big(-\frac{N-1}{4N} \tilde{z}^2\Big)
\end{equation}
an effective potential,
where the scaled momentum $\tilde{p}=\sqrt{t}p/\hbar$ has been introduced. $p$ is the conjugate momentum of $x$.
$V_N(\tilde{p})$ is a product of a polynomial in $(\sqrt{t}\tilde{p})^2$ of degree $(N-1)$ (originating from the fermionic nature of the $1$-RDM) and a Gaussian $\exp{[-\frac{N}{N-1}(\sqrt{t}\tilde{p})^2]}$ (see Appendix \ref{app:FandV}).
\begin{figure}
\centering
\includegraphics [scale=0.40]{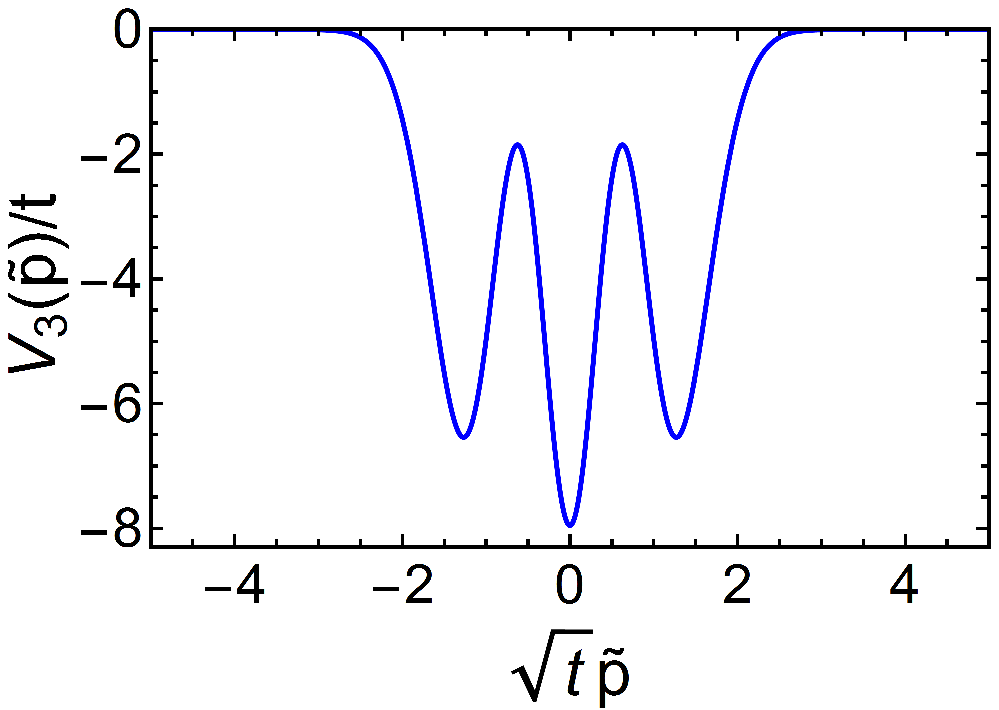}
\hspace{0.2cm}
\includegraphics [scale=0.40]{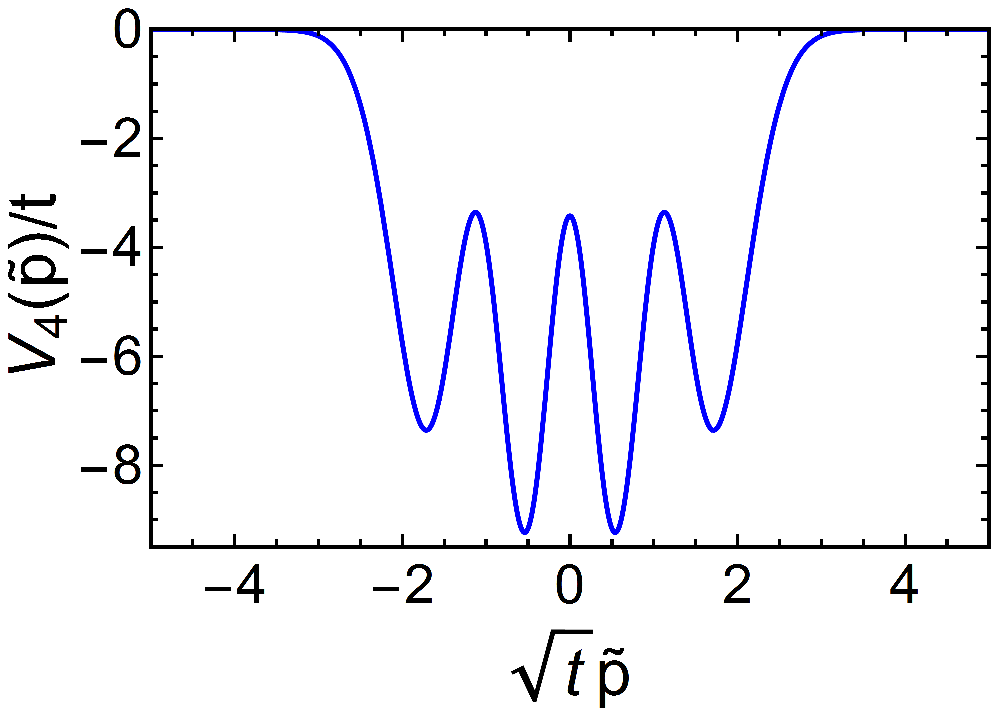}
\caption{\label{fig1} (Color online) Scaled effective potential as a function of $\sqrt{t}\tilde{p}$ for $N=3$ (left panel) and  $N=4$ (right panel).}
\end{figure}

Figure~\ref{fig1} depicts $V_N(\tilde{p})$ for $N=3$ and $N=4$. One observes that its number of minima equals $N$.
Since $V_N(\tilde{p})$ for arbitrary $N$ is symmetric, one of the minima must be at $\tilde{p}=0$, for $N$ odd. In this case it is the \textit{global} minimum, which we checked systematically up to $N=19$ and for some exemplary $N$ up to $N=101$. For $N$ even, the global minimum is twofold degenerate, which we checked for $N=2, 4, 6, \ldots, 20$ and for some exemplary $N$ up to $N=100$. There is little doubt that these properties  hold for all $N$. This qualitatively different behavior of  $V_N(\tilde{p})$ implies that the  spectrum $\{\lambda_k\}$ of the ground state $1$-RDM will qualitatively differ for an odd and even number of particles. Without solving the eigenvalue equation (\ref{eq11}) one can already  predict that the  spectrum $\sigma$ of the 1-RDM consists of two parts $\sigma^{(up)}$ for $k<k_{*}(t)$ and  $\sigma^{(low)}$ for $k>k_{*}(t)$ where $k_{*}(t)  \sim 1/t $ is the $k$-value for which $(-\lambda_k)$ equals the height of the highest maximum of $V_N(\tilde{p})$. $\sigma^{(low)}_{ent}$ consists of ``isolated'' eigenvalues, only, and does not depend qualitatively on the number parity.  In contrast, the upper part, $\sigma^{(up)}_{ent}$, differs qualitatively for $N$ odd and even. For $N$ odd it consists of subsets with ``isolated'' eigenvalues and subsets of pairs of quasidegenerate eigenvalues, whereas for  $N$ even only subsets with quasidegenerate eigenvalues occur.

This number parity effect can be illustrated by calculating analytically the \textit{largest} eigenvalues of the $1$-RDM corresponding to the \textit{low-lying} eigenvalues $(-\lambda_k)$ of the `particle' in the effective potential. This will be done by use of the harmonic approximation $V_N(\tilde{p}) \simeq V_{N}^{(min)}+ \frac{1}{2}V''^{(min)}_{N}(\tilde{p}-\tilde{p}_{min})^2$ for the global minimum. Then, Eq.~(\ref{eq11}) reduces to the eigenvalue equation of a harmonic oscillator with mass $\tilde{m}$, frequency $\tilde{\Omega}=\sqrt{V''^{(min)}_{N}/\tilde{m}}$ and eigenvalues $\bar{\lambda}_k= -(\lambda_k+V_{N}^{(min)})=\hbar \tilde{\Omega}(k-\frac{1}{2})$.
Accordingly, we obtain
\begin{equation} \label{eq14}
\lambda_k \simeq t \ \alpha
\Big[1- t \ \beta \Big(k-\frac{1}{2}\Big)  \Big]
\end{equation}
for $k=1,2,3,\ldots$ .  The coefficients follow from
\begin{equation} \label{eq14a}
\alpha=-V^{(min)}_{N}/t  \,,\quad  \beta =\sqrt{2N  V''^{(min)}_{N}/(-tV^{(min)}_N)}\,,
\end{equation}
which are of $O(t^{0})$.

For $N$ \textit{odd}, the global minimum of $V_N(\tilde{p})$ is nondegenerate at $\tilde{p}=0$.  Hence the smallest eigenvalues ($-\lambda_{k}$) and therefore the largest eigenvalues $\lambda_{k}$ of Eq.~(\ref{eq10}) are ``isolated''. In case of $N$ \textit{even}, the global minimum is twofold degenerate such that the tunneling between both minima leads to a splitting $(\Delta \lambda_{k}^{+}+\Delta\lambda_{k}^{-})$. Therefore, the eigenvalues appear in quasidegenerate pairs \{$\lambda_{2k-1},\lambda_{2k}$\}  given by
\begin{equation} \label{eq15}
\lambda_{2k-1} \simeq t \ \alpha
\Big[1- t \ \beta \Big(k-\frac{1}{2}\Big)  \Big]+\Delta\lambda_{k}^{+}
\end{equation}
for $k=1,2,3,\ldots$ \ . $\lambda_{2k}$ follows from Eq.~(\ref{eq15}) by replacing $\Delta\lambda_{k}^{+}$ by $(-\Delta\lambda_{k}^{-})$.

The results (\ref{eq14}) and (\ref{eq15}) show that  $\lambda_{k}$ is of $O(t)$. Its $k$-dependency is proportional to $t^2$. Since these eigenvalues are the occupancies of the $1$-particle states $\chi_{k}(x)$ they are nonnegative.  Hence, the validity of Eqs.~(\ref{eq14}), (\ref{eq15}) is limited to
$k < k_{*} \sim 1/t$. They are also based on the harmonic approximation, which restricts their validity even more(see below). Furthermore, the tunneling splitting $(\Delta\lambda_{k}^{+}+\Delta\lambda_{k}^{-})$ can be estimated.  Equation (\ref{eq13})  shows that $V_N(\tilde{p})=O(t)$ and that it  varies on a scale $~1/\sqrt{t}$. This implies $\tilde{m}=O(t^{-2})$, a potential barrier $\Delta V_N=O(t)$
and a tunneling distance $\Delta \tilde{p}=O(1/\sqrt{t})$. Using that the splitting is proportional to $\exp{(-\sqrt{\tilde{m}\Delta V_N}\Delta \tilde{p})/\hbar)}$ it follows that
$(\Delta\lambda_{k}^{+}+\Delta\lambda_{k}^{-}) \sim \exp{(-O(1/t))}$, i.e.~for $N$ even the pairs ($\lambda_{2k-1}, \lambda_{2k})$  become  quasidegenerate in the regime of strong coupling.

In order to test these predictions we have determined the eigenvalues by solving numerically
the original eigenvalue equation (\ref{eq10}) for $N=3$ up to $N=8$. Note that $N=2$ is a special case, since there, all eigenvalues are automatically twofold degenerate as a hamiltonian-independent consequence of the fermionic exchange statistics (see Theorem 4.1. in Ref.~\cite{Col2}).
\begin{figure}
\centering
\includegraphics [scale=0.60]{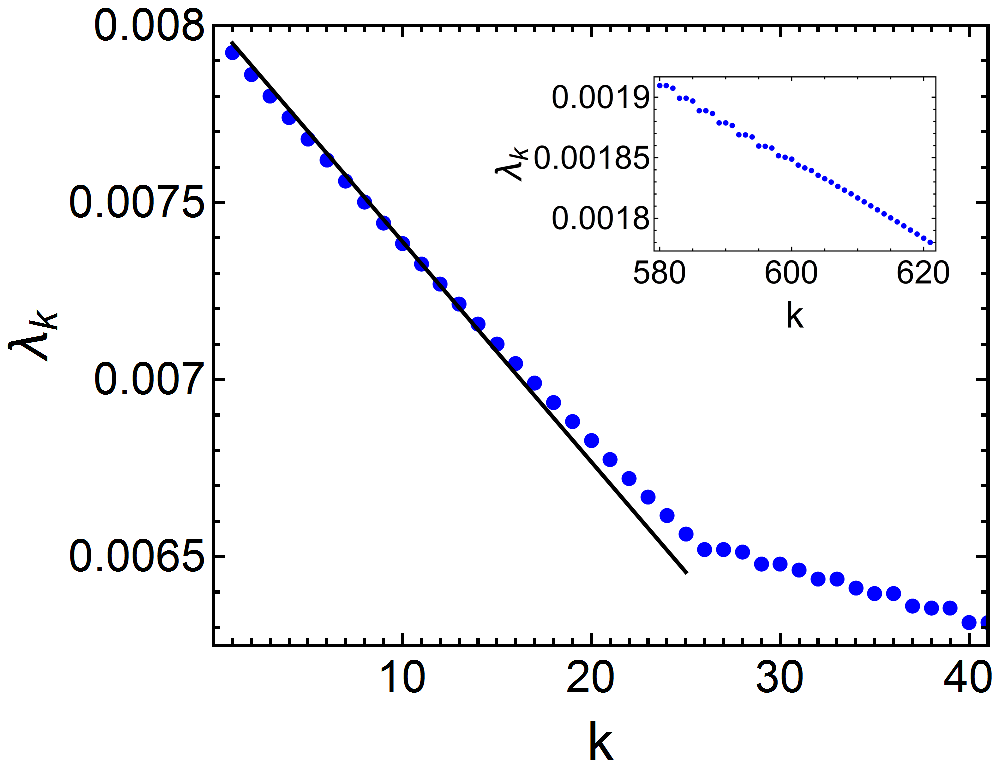}
\includegraphics [scale=0.60]{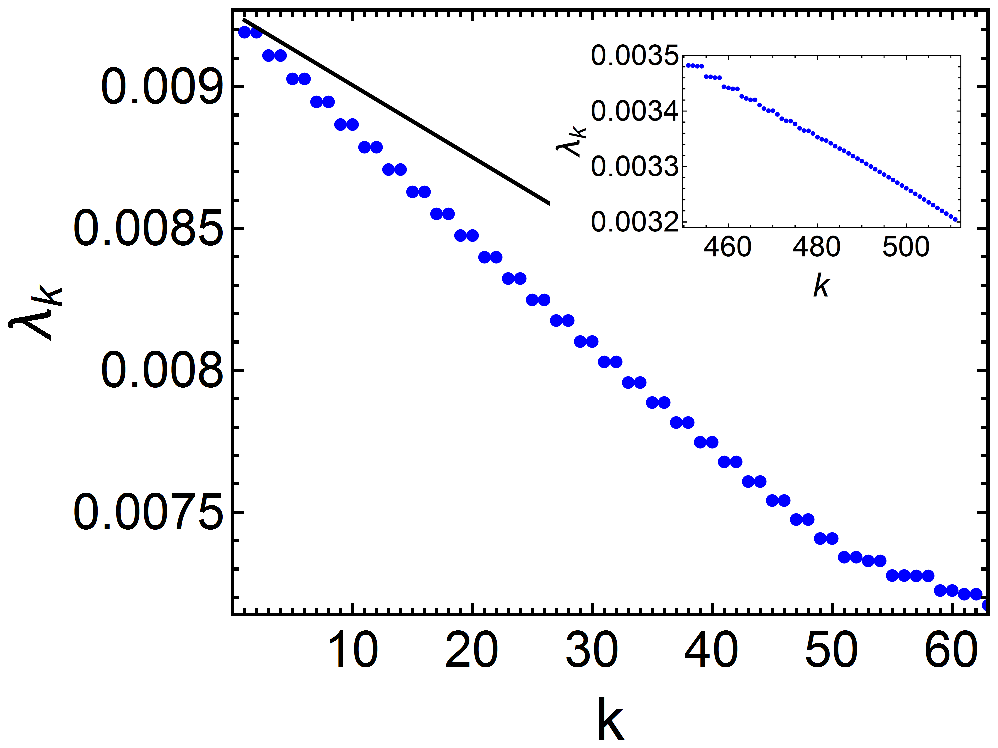}
\caption{\label{fig2} (Color online)  Spectrum $\{\lambda_{k}\}$ for $t=10^{-3}$.
 Upper panel: \, $N=3$ and $1 \leq k \leq 40 $,  lower panel: \, $N=4$ and $1 \leq k \leq 62 $. The solid lines present the corresponding analytical result for $\{\lambda_{k}\}$ from Eq.~(\ref{eq14}) and from the first term of Eq.~(\ref{eq15}), respectively. The insets show the final crossover from the regime of quasidegenerate to that of non-quasidegenerate eigenvalues.}
\end{figure}
Figure \ref{fig2} presents the larger eigenvalues for the coupling $t=10^{-3}$ for $N=3$ and $N=4$. In case of $N=3$ the subsequent eigenvalues have almost the same distance which is of $O(t^2)$ whereas for $N=4$ they occur in quasidegenerate pairs with $\lambda_{2k-1}-\lambda_{2k+1}=O(t^2)$ and  $\lambda_{2k-1}-\lambda_{2k} \sim \exp{(-O(1/t))}$, as predicted by our analysis above. Comparing for $N=3$ the numerical result  with the analytical one, Eq.~(\ref{eq14}), shows very good agreement for $\lambda_1$ and the slope $d\lambda_k/dk$, for small $k$. For $N=4$,  $\lambda_1$ is also well reproduced. However, the analytical and numerical values for the slope $d\lambda_k/dk$ deviate stronger from each other. This is a consequence of the fact that
in contrast to $N=3$ the harmonic approximation is rather poor due to $(\tilde{p}-\tilde{p}_{min})^3$-anharmonicities of $V_4$ around $\tilde{p}_{min}$. Note, our major achievement is not result (\ref{eq14}) and (\ref{eq15}) for the largest eigenvalues, but
\begin{enumerate}[label=(\roman*)]
\item  the qualitatively different  spectrum of the 1-RDM for $N$ odd and even in the strong coupling limit
\item the generation and annihilation of quasidegenerate pairs of eigenvalues for those indices $k$ for which $(-\lambda_k)$ becomes equal to the height of corresponding minima and maxima, respectively, of $V_N(\tilde{p})$
\end{enumerate}

\begin{figure}
\centering
\includegraphics [scale=0.40]{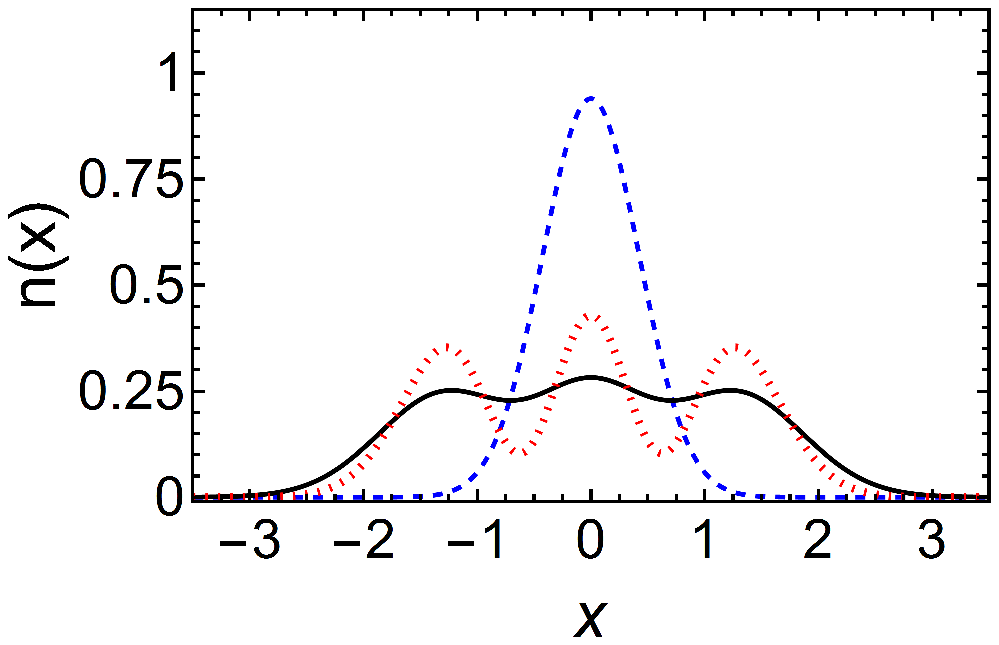}
\hspace{0.2cm}
\includegraphics [scale=0.40]{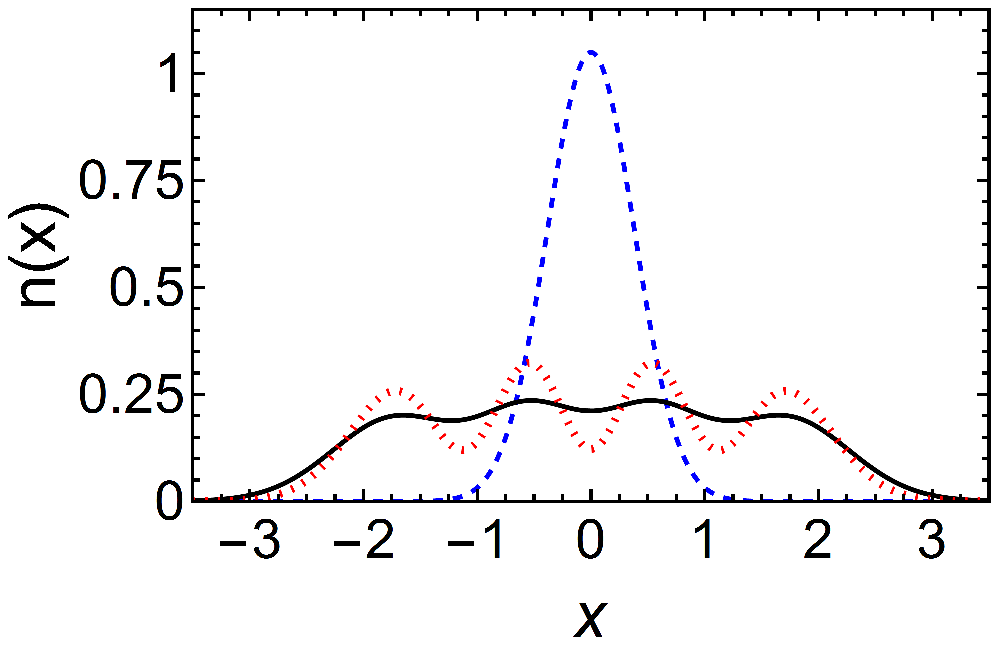}
\caption{\label{fig3}(Color online) $x$-dependency of the $1$-particle density for $N=3$ (left panel) and $N=4$ (right panel), where we set $l_-\equiv 1$.  Blue dashed,  black solid and  red dotted line correspond to  $t=0.1$ (attractive), $t=1$ (free fermions), and   $t=10$ (repulsive), respectively (for $t=10$ it is plotted $10\,n(10x)$). }
\end{figure}

Figure \ref{fig2} demonstrates the creation of pairs of quasidegenerate eigenvalues and its inset illustrates their annihilation, e.g.~at the highest maximum of $V_N(\tilde{p})$, where the crossover to ``isolated'' eigenvalues of $\sigma^{(low)}_{ent}$ occurs.

Since under an increase of $N$, $V_N(\tilde{p})$ develops more and more extrema,  there will be more and more regimes  with groups of a different number of quasidegenerate pairs.
In order to resolve these regimes for large $N$, $t$ must become small enough. Since the $t$ and $N$ dependence  occurs as $tN$  (see Appendix \ref{app:FandV}) it must be $t \ll t_{*}(N) \sim 1/N$. For macroscopically large $N$ of $\mathcal{O}(10^{23})$ this requires values for $t$ which are not realizable in  experiments. Yet, for $N$ of $\mathcal{O}(10^{2})$, for which already macroscopic properties are present, this should be feasible.
Accordingly, \textit{few-fermion} systems
are particularly suitable to observe these qualitative features of the  spectrum of the 1-RDM.

$\lambda_{k}$ for $k \gg k_{*} \sim 1/t$ can be determined analytically.
With the approach discussed in Ref.~\cite{CS2013NO} we obtain (see Appendix \ref{app:NONs})
\begin{equation} \label{eq16}
\lambda_k \sim  (k t)^{N-1} \exp [-\frac{2N}{\sqrt{N-1}} \ t \ (k-\frac{1}{2})] \ ,\
\end{equation}
\textit{independent} of the parity of $N$.

\begin{figure}
\centering
\includegraphics [scale=0.40]{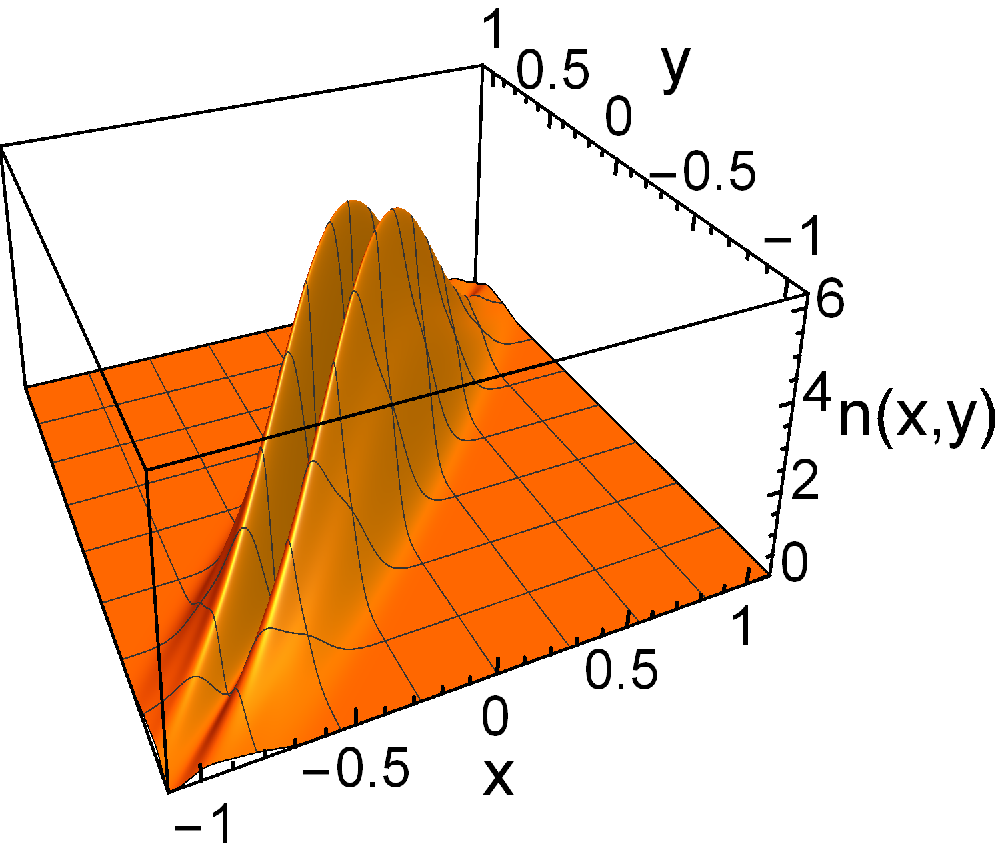}
\hspace{0.2cm}
\includegraphics [scale=0.40]{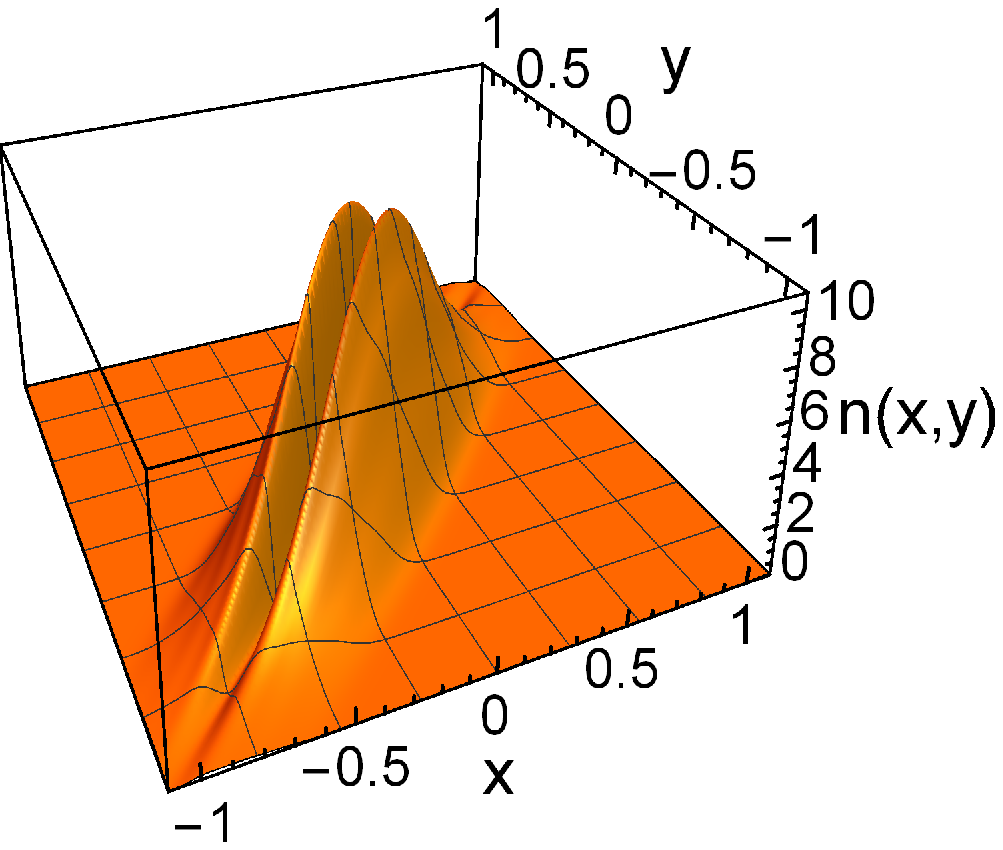}
\caption{\label{fig4}(Color online) $x$- and $y$-dependency of the $2$-particle density  $n(x,y)$ (not normalized and we set $l_-\equiv 1$) for $t=0.1$,  $N=3$ (left panel) and $N=4$ (right panel).}
\end{figure}

We have also calculated analytically the 1-particle and 2-particle density $n(x)\equiv\rho_1(x;x)$ and  $n(x,y)\equiv\rho_2(x,y;x,y)$, respectively. The latter
follows from the 2-RDM $\rho_2(x,y;x',y')$. $n(x)$ is shown in Figure \ref{fig3} for $N=3$ and $N=4$.  For noninteracting fermions (black solid line), the `layering' of the particles  within the harmonic trap can be seen. This ``shell structure'' also exists for repulsive (red dotted line) and weak attractive coupling, but disappears completely for strong attractive interactions (blue dashed line), becoming qualitatively independent of $N$, which we checked up to $N=20$. The duality discussed in Ref.~\cite{duality} implies that the $1$-particle density in momentum space for strong \emph{repulsive} interaction behaves similarly as $n(x)$ for strong \emph{attractive} interactions, i.e.~it becomes structureless, as well.
Quite similar behavior has been found for $n(x,y)$.   As demonstrated by Figure \ref{fig4} the `layering'(not shown) for  $n(x,y)$ disappears again for strong attractive coupling, and does not exhibits any qualitative sensitivity on $N$ (cf.~left and right panel of Figure \ref{fig4}). All these properties also hold for the correlation function $C(x,y)=n(x,y)/n(x)n(y)$.




\paragraph{Summary and Conclusions.---}
We have shown that the $1$-particle description in form of the $1$-RDM exhibits an odd-even effect. The  spectrum of the 1-RDM related to the fermionic ground state of our one-dimensional harmonic system differs qualitatively for an even and an odd number of particles.
This effect does only occur for \textit{strong} attractive and repulsive (due to the duality property \cite{duality}) interactions.
The number-parity effect does not exist for bosons. Therefore it must originate from the repulsive nature implied by  the fermionic exchange statistics (antisymmetry of the wave function). One may wonder how far the interplay between \textit{strong} pair interactions (particularly for attractive ones) and the exchange symmetry leads to new phenomena for fermions, beyond the present parity effect.  Also the investigation of the existence of the odd-even effect in more than one spatial dimension will be of interest.

It would also be  interesting to develop tools which make it  possible to investigate these predictions by experiments. In that case the parameter $t$ has to be tuned (see below) such that the splitting of the quasidegenerate
eigenvalues is still large enough in order to be resolved. Since our model involves harmonic pair interactions one might be tempted to deny the relevance of our findings for realistic systems. There are two reasons why this might be not true. \textit{First}, for arbitrary pair potential the particles in a trap form a one-dimensional lattice. Expanding the potential up to quadratic terms in the displacements with respect to the classical groundstate  will result in a harmonic model similar to that studied by us. This has recently been done for a one-dimensional $N$-particle system with long-range inverse power-law potential in order to calculate the von Neumann entanglement entropy \cite{koscik2015neumann}. \textit{Second}, and even more important, the specific form of the pair interactions may not be as important as one might believe. As pointed out above the odd-even effect for the spectrum of the fermionic 1-RDM is a result of the polynomial prefactor in Eq.~(\ref{eq2}) which makes the wave function totally antisymmetric. This requirement of antisymmetry has also been the guide leading to Laughlin's wave function for describing the fractional quantum Hall effect. This wave function again has a preexponential factor $\Big[\prod_{1 \leq i < j \leq N} (z_i - z_j)^{p}\Big]$, with $p$ an odd natural number and $z_i$ the complex variable specifying the position of the i-th particle in the plane. Laughlin's ansatz is a surprisingly good approximation of the two-dimensional electron system's ground state, not only for a Coulombic pair potential but for harmonic interactions, as well \cite{Laughlin1983}. This suggests that the specific form of the pair interactions is less important which is supported by the exact ground state solution of the one-dimensional Calogero-Sutherland model \cite{CS1,CS2,CS3,CS4,CS5}. Besides a harmonic trap potential this model contains pair interactions $g(x_i - x_j)^{-2}$. Its ground state involves a preexponential factor  $\Big[\prod_{1 \leq i < j \leq N} (x_i - x_j)^{p}\Big]$ with $p(g)=\frac{1}{2} (1 \pm \sqrt{1+4g})$. Choosing the coupling constant such that $p(g)$ is an odd natural number one obtains the ground state for N fermions.

Therefore our results may also hold for ultracold fermionic
atoms in an optical trap  interacting by a  contact potential \cite{ultracoldBook1,BlochReview,Feshbach}.  The interaction can be tuned from attractive to repulsive. In particular it can be made arbitrary strong by approaching the Feshbach resonance. This would allow to experimentally approach  the strong coupling limit discussed in the present paper . Of course, another possibility to realize that limit is the decrease of the trap frequency. For instance for $^{7}Li$  the trap frequency in the experimental setup in Ref. \cite{PauliPressure} is about thirty times smaller
than in the setup of  Ref.~\cite{Jochim2species}.

The odd-even effect may also initiate and guide a new direction in density and reduced-density-matrix-functional theory.
Although three decades ago, it had been argued that the
$N$-dependency of the ground state energy implies an $N$-dependency of the density functionals $\mathcal{F}_{N}$ \cite{LiebDFT}, all of the prominent functionals used today do not exhibit an explicit dependency on $N$ (see e.g.~\cite{DFTbook1}). The parity effect found by us is the first  demonstration of a subtle $N$-dependency of the spectrum of the ground state's $1$-RDM, and therefore also of $\mathcal{F}_{N}$. In addition, the fact that the 1-particle density $n(x)$ in the strong coupling regime does not show any sensitivity on the number parity suggests that the $N$ dependence within reduced-density-matrix-functional theory \cite{Gilbert,Col3} is much more subtle than in ordinary density-functional theory \cite{Hohenberg}.

\paragraph*{Acknowledgements.---}
We would like to thank D.~Jaksch, S. Jochim, F.~Schmidt-Kaler, P.~van Dongen and V.~Vedral for helpful discussions. We are particularly grateful to P.~van Dongen for numerous valuable comments on our manuscript.
CS gratefully acknowledges financial support from the Swiss National Science Foundation (Grant P2EZP2 152190) and from the Oxford Martin Programme on Bio-Inspired Quantum Technologies.

\bibliography{bibliography}

\begin{thebibliography}{45}%
\makeatletter
\providecommand \@ifxundefined [1]{%
 \@ifx{#1\undefined}
}%
\providecommand \@ifnum [1]{%
 \ifnum #1\expandafter \@firstoftwo
 \else \expandafter \@secondoftwo
 \fi
}%
\providecommand \@ifx [1]{%
 \ifx #1\expandafter \@firstoftwo
 \else \expandafter \@secondoftwo
 \fi
}%
\providecommand \natexlab [1]{#1}%
\providecommand \enquote  [1]{``#1''}%
\providecommand \bibnamefont  [1]{#1}%
\providecommand \bibfnamefont [1]{#1}%
\providecommand \citenamefont [1]{#1}%
\providecommand \href@noop [0]{\@secondoftwo}%
\providecommand \href [0]{\begingroup \@sanitize@url \@href}%
\providecommand \@href[1]{\@@startlink{#1}\@@href}%
\providecommand \@@href[1]{\endgroup#1\@@endlink}%
\providecommand \@sanitize@url [0]{\catcode `\\12\catcode `\$12\catcode
  `\&12\catcode `\#12\catcode `\^12\catcode `\_12\catcode `\%12\relax}%
\providecommand \@@startlink[1]{}%
\providecommand \@@endlink[0]{}%
\providecommand \url  [0]{\begingroup\@sanitize@url \@url }%
\providecommand \@url [1]{\endgroup\@href {#1}{\urlprefix }}%
\providecommand \urlprefix  [0]{URL }%
\providecommand \Eprint [0]{\href }%
\providecommand \doibase [0]{http://dx.doi.org/}%
\providecommand \selectlanguage [0]{\@gobble}%
\providecommand \bibinfo  [0]{\@secondoftwo}%
\providecommand \bibfield  [0]{\@secondoftwo}%
\providecommand \translation [1]{[#1]}%
\providecommand \BibitemOpen [0]{}%
\providecommand \bibitemStop [0]{}%
\providecommand \bibitemNoStop [0]{.\EOS\space}%
\providecommand \EOS [0]{\spacefactor3000\relax}%
\providecommand \BibitemShut  [1]{\csname bibitem#1\endcsname}%
\let\auto@bib@innerbib\@empty
\bibitem [{\citenamefont {Kramers}(1930)}]{Kramers}%
  \BibitemOpen
  \bibfield  {author} {\bibinfo {author} {\bibfnamefont {H.A.}\ \bibnamefont
  {Kramers}},\ }\bibfield  {title} {\enquote {\bibinfo {title} {Th\'eorie
  g\'en\'erale de la rotation paramagn\'etique dans les cristaux},}\
  }\href@noop {} {\bibfield  {journal} {\bibinfo  {journal} {Proc. Amsterdam
  Acad.}\ }\textbf {\bibinfo {volume} {33}},\ \bibinfo {pages} {959} (\bibinfo
  {year} {1930})}\BibitemShut {NoStop}%
\bibitem [{\citenamefont {Haldane}(1983)}]{Haldaneparity1}%
  \BibitemOpen
  \bibfield  {author} {\bibinfo {author} {\bibfnamefont {F.D.M.}\ \bibnamefont
  {Haldane}},\ }\bibfield  {title} {\enquote {\bibinfo {title} {Continuum
  dynamics of the 1-d {H}eisenberg antiferromagnet: Identification with the
  {O}(3) nonlinear sigma model},}\ }\href
  {http://www.sciencedirect.com/science/article/pii/037596018390631X}
  {\bibfield  {journal} {\bibinfo  {journal} {Phys. Lett. A}\ }\textbf
  {\bibinfo {volume} {93}},\ \bibinfo {pages} {464 -- 468} (\bibinfo {year}
  {1983})}\BibitemShut {NoStop}%
\bibitem [{\citenamefont {Z\"urn}\ \emph {et~al.}(2013)\citenamefont {Z\"urn},
  \citenamefont {Wenz}, \citenamefont {Murmann}, \citenamefont {Bergschneider},
  \citenamefont {Lompe},\ and\ \citenamefont {Jochim}}]{Jochimpairing}%
  \BibitemOpen
  \bibfield  {author} {\bibinfo {author} {\bibfnamefont {G.}~\bibnamefont
  {Z\"urn}}, \bibinfo {author} {\bibfnamefont {A.~N.}\ \bibnamefont {Wenz}},
  \bibinfo {author} {\bibfnamefont {S.}~\bibnamefont {Murmann}}, \bibinfo
  {author} {\bibfnamefont {A.}~\bibnamefont {Bergschneider}}, \bibinfo {author}
  {\bibfnamefont {T.}~\bibnamefont {Lompe}}, \ and\ \bibinfo {author}
  {\bibfnamefont {S.}~\bibnamefont {Jochim}},\ }\bibfield  {title} {\enquote
  {\bibinfo {title} {Pairing in few-fermion systems with attractive
  interactions},}\ }\href {\doibase 10.1103/PhysRevLett.111.175302} {\bibfield
  {journal} {\bibinfo  {journal} {Phys. Rev. Lett.}\ }\textbf {\bibinfo
  {volume} {111}},\ \bibinfo {pages} {175302} (\bibinfo {year}
  {2013})}\BibitemShut {NoStop}%
\bibitem [{\citenamefont {Smith}(1966)}]{Smith1966}%
  \BibitemOpen
  \bibfield  {author} {\bibinfo {author} {\bibfnamefont {D.~W.}\ \bibnamefont
  {Smith}},\ }\bibfield  {title} {\enquote {\bibinfo {title}
  {$n$-representability problem for fermion density matrices. ii. the
  first-order density matrix with $n$ even},}\ }\href {\doibase
  10.1103/PhysRev.147.896} {\bibfield  {journal} {\bibinfo  {journal} {Phys.
  Rev.}\ }\textbf {\bibinfo {volume} {147}},\ \bibinfo {pages} {896--898}
  (\bibinfo {year} {1966})}\BibitemShut {NoStop}%
\bibitem [{\citenamefont {Coleman}(1963)}]{Col2}%
  \BibitemOpen
  \bibfield  {author} {\bibinfo {author} {\bibfnamefont {A.~J.}\ \bibnamefont
  {Coleman}},\ }\bibfield  {title} {\enquote {\bibinfo {title} {Structure of
  fermion density matrices},}\ }\href {\doibase 10.1103/RevModPhys.35.668}
  {\bibfield  {journal} {\bibinfo  {journal} {Rev. Mod. Phys.}\ }\textbf
  {\bibinfo {volume} {35}},\ \bibinfo {pages} {668--686} (\bibinfo {year}
  {1963})}\BibitemShut {NoStop}%
\bibitem [{\citenamefont {{Klyachko}}(2004)}]{Kly4}%
  \BibitemOpen
  \bibfield  {author} {\bibinfo {author} {\bibfnamefont {A.}~\bibnamefont
  {{Klyachko}}},\ }\bibfield  {title} {\enquote {\bibinfo {title} {{Quantum
  marginal problem and representations of the symmetric group}},}\ }\href
  {http://arxiv.org/abs/quant-ph/0409113} {\bibfield  {journal} {\bibinfo
  {journal} {arXiv:0409113}\ } (\bibinfo {year} {2004})}\BibitemShut {NoStop}%
\bibitem [{\citenamefont {Daftuar}\ and\ \citenamefont {Hayden}(2005)}]{Daft}%
  \BibitemOpen
  \bibfield  {author} {\bibinfo {author} {\bibfnamefont {S.}~\bibnamefont
  {Daftuar}}\ and\ \bibinfo {author} {\bibfnamefont {P.}~\bibnamefont
  {Hayden}},\ }\bibfield  {title} {\enquote {\bibinfo {title} {Quantum state
  transformations and the {S}chubert calculus},}\ }\href
  {http://www.sciencedirect.com/science/article/pii/S0003491604001770}
  {\bibfield  {journal} {\bibinfo  {journal} {Annals Phys.}\ }\textbf {\bibinfo
  {volume} {315}},\ \bibinfo {pages} {80 -- 122} (\bibinfo {year}
  {2005})}\BibitemShut {NoStop}%
\bibitem [{\citenamefont {Klyachko}(2006)}]{Kly2}%
  \BibitemOpen
  \bibfield  {author} {\bibinfo {author} {\bibfnamefont {A.}~\bibnamefont
  {Klyachko}},\ }\bibfield  {title} {\enquote {\bibinfo {title} {Quantum
  marginal problem and n-representability},}\ }\href
  {http://stacks.iop.org/1742-6596/36/i=1/a=014} {\bibfield  {journal}
  {\bibinfo  {journal} {J. Phys.: Conf. Series}\ }\textbf {\bibinfo {volume}
  {36}},\ \bibinfo {pages} {72} (\bibinfo {year} {2006})}\BibitemShut {NoStop}%
\bibitem [{\citenamefont {Christandl}\ and\ \citenamefont
  {Mitchison}(2006)}]{MC2006}%
  \BibitemOpen
  \bibfield  {author} {\bibinfo {author} {\bibfnamefont {M.}~\bibnamefont
  {Christandl}}\ and\ \bibinfo {author} {\bibfnamefont {G.}~\bibnamefont
  {Mitchison}},\ }\bibfield  {title} {\enquote {\bibinfo {title} {The spectra
  of quantum states and the {K}ronecker coefficients of the symmetric group},}\
  }\href {\doibase 10.1007/s00220-005-1435-1} {\bibfield  {journal} {\bibinfo
  {journal} {Commun. Math. Phys.}\ }\textbf {\bibinfo {volume} {261}},\
  \bibinfo {pages} {789--797} (\bibinfo {year} {2006})}\BibitemShut {NoStop}%
\bibitem [{\citenamefont {Altunbulak}\ and\ \citenamefont
  {Klyachko}(2008)}]{Kly3}%
  \BibitemOpen
  \bibfield  {author} {\bibinfo {author} {\bibfnamefont {M.}~\bibnamefont
  {Altunbulak}}\ and\ \bibinfo {author} {\bibfnamefont {A.}~\bibnamefont
  {Klyachko}},\ }\bibfield  {title} {\enquote {\bibinfo {title} {The {P}auli
  principle revisited},}\ }\href {\doibase 10.1007/s00220-008-0552-z}
  {\bibfield  {journal} {\bibinfo  {journal} {Commun. Math. Phys.}\ }\textbf
  {\bibinfo {volume} {282}},\ \bibinfo {pages} {287--322} (\bibinfo {year}
  {2008})}\BibitemShut {NoStop}%
\bibitem [{\citenamefont {Altunbulak}(2008)}]{Altun}%
  \BibitemOpen
  \bibfield  {author} {\bibinfo {author} {\bibfnamefont {M.}~\bibnamefont
  {Altunbulak}},\ }\emph {\bibinfo {title} {The {P}auli principle,
  representation theory, and geometry of flag varieties}},\ \href
  {http://www.thesis.bilkent.edu.tr/0003572} {Ph.D. thesis},\ \bibinfo
  {school} {Bilkent University} (\bibinfo {year} {2008})\BibitemShut {NoStop}%
\bibitem [{\citenamefont {Schilling}(2014{\natexlab{a}})}]{CSQMath12}%
  \BibitemOpen
  \bibfield  {author} {\bibinfo {author} {\bibfnamefont {C.}~\bibnamefont
  {Schilling}},\ }\enquote {\bibinfo {title} {The quantum marginal problem},}\
  in\ \href {\doibase 10.1142/9789814618144_0010} {\emph {\bibinfo {booktitle}
  {Mathematical Results in Quantum Mechanics}}}\ (\bibinfo  {publisher} {World
  Scientific},\ \bibinfo {year} {2014})\ Chap.~\bibinfo {chapter} {-1}, pp.\
  \bibinfo {pages} {165--176}\BibitemShut {NoStop}%
\bibitem [{\citenamefont {Schilling}(2014{\natexlab{b}})}]{CSthesis}%
  \BibitemOpen
  \bibfield  {author} {\bibinfo {author} {\bibfnamefont {C.}~\bibnamefont
  {Schilling}},\ }\emph {\bibinfo {title} {Quantum marginal problem and its
  physical relevance}},\ \href {\doibase 10.3929/ethz-a-010139282} {Ph.D.
  thesis},\ \bibinfo  {school} {ETH-Z\"urich} (\bibinfo {year}
  {2014}{\natexlab{b}})\BibitemShut {NoStop}%
\bibitem [{\citenamefont {Lopes}(2015)}]{Alex}%
  \BibitemOpen
  \bibfield  {author} {\bibinfo {author} {\bibfnamefont {A.}~\bibnamefont
  {Lopes}},\ }\emph {\bibinfo {title} {Pure univariate quantum marginals and
  electronic transport properties of geometrically frustrated systems}},\ \href
  {https://www.freidok.uni-freiburg.de/fedora/objects/freidok:10057/datastreams/FILE1/content}
  {Ph.D. thesis},\ \bibinfo  {school} {University of Freiburg} (\bibinfo {year}
  {2015})\BibitemShut {NoStop}%
\bibitem [{\citenamefont {Davidson}(1976)}]{Dav}%
  \BibitemOpen
  \bibfield  {author} {\bibinfo {author} {\bibfnamefont {E.R.}\ \bibnamefont
  {Davidson}},\ }\href@noop {} {\emph {\bibinfo {title} {Reduced Density
  Matrices in Quantum Chemistry}}}\ (\bibinfo  {publisher} {Academic Press, New
  York},\ \bibinfo {year} {1976})\BibitemShut {NoStop}%
\bibitem [{\citenamefont {Liu}\ \emph {et~al.}(2007)\citenamefont {Liu},
  \citenamefont {Christandl},\ and\ \citenamefont {Verstraete}}]{QMA}%
  \BibitemOpen
  \bibfield  {author} {\bibinfo {author} {\bibfnamefont {Y.-K.}\ \bibnamefont
  {Liu}}, \bibinfo {author} {\bibfnamefont {M.}~\bibnamefont {Christandl}}, \
  and\ \bibinfo {author} {\bibfnamefont {F.}~\bibnamefont {Verstraete}},\
  }\bibfield  {title} {\enquote {\bibinfo {title} {Quantum computational
  complexity of the $n$-representability problem: {QMA} complete},}\ }\href
  {\doibase 10.1103/PhysRevLett.98.110503} {\bibfield  {journal} {\bibinfo
  {journal} {Phys. Rev. Lett.}\ }\textbf {\bibinfo {volume} {98}},\ \bibinfo
  {pages} {110503} (\bibinfo {year} {2007})}\BibitemShut {NoStop}%
\bibitem [{\citenamefont {Gilbert}(1975)}]{Gilbert}%
  \BibitemOpen
  \bibfield  {author} {\bibinfo {author} {\bibfnamefont {T.~L.}\ \bibnamefont
  {Gilbert}},\ }\bibfield  {title} {\enquote {\bibinfo {title}
  {Hohenberg-{K}ohn theorem for nonlocal external potentials},}\ }\href
  {\doibase 10.1103/PhysRevB.12.2111} {\bibfield  {journal} {\bibinfo
  {journal} {Phys. Rev. B}\ }\textbf {\bibinfo {volume} {12}},\ \bibinfo
  {pages} {2111--2120} (\bibinfo {year} {1975})}\BibitemShut {NoStop}%
\bibitem [{\citenamefont {Erdahl}\ and\ \citenamefont {Smith}(1987)}]{Col3}%
  \BibitemOpen
  \bibfield  {author} {\bibinfo {author} {\bibfnamefont {R.M.}\ \bibnamefont
  {Erdahl}}\ and\ \bibinfo {author} {\bibfnamefont {V.H.}\ \bibnamefont
  {Smith}},\ }\href@noop {} {\emph {\bibinfo {title} {Density matrices and
  density functionals}}}\ (\bibinfo  {publisher} {D. Reidel Publishing Company,
  Dordrecht},\ \bibinfo {year} {1987})\BibitemShut {NoStop}%
\bibitem [{\citenamefont {Hohenberg}\ and\ \citenamefont
  {Kohn}(1964)}]{Hohenberg}%
  \BibitemOpen
  \bibfield  {author} {\bibinfo {author} {\bibfnamefont {P.}~\bibnamefont
  {Hohenberg}}\ and\ \bibinfo {author} {\bibfnamefont {W.}~\bibnamefont
  {Kohn}},\ }\bibfield  {title} {\enquote {\bibinfo {title} {Inhomogeneous
  electron gas},}\ }\href {\doibase 10.1103/PhysRev.136.B864} {\bibfield
  {journal} {\bibinfo  {journal} {Phys. Rev.}\ }\textbf {\bibinfo {volume}
  {136}},\ \bibinfo {pages} {B864--B871} (\bibinfo {year} {1964})}\BibitemShut
  {NoStop}%
\bibitem [{\citenamefont {Johnson}\ and\ \citenamefont
  {Payne}(1991)}]{QuantumDot}%
  \BibitemOpen
  \bibfield  {author} {\bibinfo {author} {\bibfnamefont {N.~F.}\ \bibnamefont
  {Johnson}}\ and\ \bibinfo {author} {\bibfnamefont {M.~C.}\ \bibnamefont
  {Payne}},\ }\bibfield  {title} {\enquote {\bibinfo {title} {Exactly solvable
  model of interacting particles in a quantum dot},}\ }\href {\doibase
  10.1103/PhysRevLett.67.1157} {\bibfield  {journal} {\bibinfo  {journal}
  {Phys. Rev. Lett.}\ }\textbf {\bibinfo {volume} {67}},\ \bibinfo {pages}
  {1157--1160} (\bibinfo {year} {1991})}\BibitemShut {NoStop}%
\bibitem [{\citenamefont {Post}(1953)}]{HarmShells}%
  \BibitemOpen
  \bibfield  {author} {\bibinfo {author} {\bibfnamefont {H.~R.}\ \bibnamefont
  {Post}},\ }\bibfield  {title} {\enquote {\bibinfo {title} {Many-particle
  systems: Derivation of a shell model},}\ }\href
  {http://stacks.iop.org/0370-1298/66/i=7/a=411} {\bibfield  {journal}
  {\bibinfo  {journal} {Proc. Soc. A}\ }\textbf {\bibinfo {volume} {66}},\
  \bibinfo {pages} {649} (\bibinfo {year} {1953})}\BibitemShut {NoStop}%
\bibitem [{\citenamefont {Zinner}\ and\ \citenamefont
  {Jensen}(2008)}]{ZinnerNucl}%
  \BibitemOpen
  \bibfield  {author} {\bibinfo {author} {\bibfnamefont {N.~T.}\ \bibnamefont
  {Zinner}}\ and\ \bibinfo {author} {\bibfnamefont {A.~S.}\ \bibnamefont
  {Jensen}},\ }\bibfield  {title} {\enquote {\bibinfo {title} {Nuclear
  $\ensuremath{\alpha}$-particle condensates: Definitions, occurrence
  conditions, and consequences},}\ }\href {\doibase 10.1103/PhysRevC.78.041306}
  {\bibfield  {journal} {\bibinfo  {journal} {Phys. Rev. C}\ }\textbf {\bibinfo
  {volume} {78}},\ \bibinfo {pages} {041306} (\bibinfo {year}
  {2008})}\BibitemShut {NoStop}%
\bibitem [{\citenamefont {Wang}\ \emph {et~al.}(2012)\citenamefont {Wang},
  \citenamefont {Wang}, \citenamefont {Yang},\ and\ \citenamefont
  {Li}}]{harmOsc2012}%
  \BibitemOpen
  \bibfield  {author} {\bibinfo {author} {\bibfnamefont {Z.-L.}\ \bibnamefont
  {Wang}}, \bibinfo {author} {\bibfnamefont {A.M.}\ \bibnamefont {Wang}},
  \bibinfo {author} {\bibfnamefont {Y.}~\bibnamefont {Yang}}, \ and\ \bibinfo
  {author} {\bibfnamefont {X.C.}\ \bibnamefont {Li}},\ }\bibfield  {title}
  {\enquote {\bibinfo {title} {Exact eigenfunctions of n-body system with
  quadratic pair potential},}\ }\href
  {http://stacks.iop.org/0253-6102/58/i=5/a=04} {\bibfield  {journal} {\bibinfo
   {journal} {Comm. Theor. Phys.}\ }\textbf {\bibinfo {volume} {58}},\ \bibinfo
  {pages} {639} (\bibinfo {year} {2012})}\BibitemShut {NoStop}%
\bibitem [{\citenamefont {Schilling}\ \emph {et~al.}(2013)\citenamefont
  {Schilling}, \citenamefont {Gross},\ and\ \citenamefont
  {Christandl}}]{CS2013}%
  \BibitemOpen
  \bibfield  {author} {\bibinfo {author} {\bibfnamefont {C.}~\bibnamefont
  {Schilling}}, \bibinfo {author} {\bibfnamefont {D.}~\bibnamefont {Gross}}, \
  and\ \bibinfo {author} {\bibfnamefont {M.}~\bibnamefont {Christandl}},\
  }\bibfield  {title} {\enquote {\bibinfo {title} {Pinning of fermionic
  occupation numbers},}\ }\href {\doibase 10.1103/PhysRevLett.110.040404}
  {\bibfield  {journal} {\bibinfo  {journal} {Phys. Rev. Lett.}\ }\textbf
  {\bibinfo {volume} {110}},\ \bibinfo {pages} {040404} (\bibinfo {year}
  {2013})}\BibitemShut {NoStop}%
\bibitem [{\citenamefont {Schilling}(2013)}]{CS2013NO}%
  \BibitemOpen
  \bibfield  {author} {\bibinfo {author} {\bibfnamefont {C.}~\bibnamefont
  {Schilling}},\ }\bibfield  {title} {\enquote {\bibinfo {title} {Natural
  orbitals and occupation numbers for harmonium: Fermions versus bosons},}\
  }\href {\doibase 10.1103/PhysRevA.88.042105} {\bibfield  {journal} {\bibinfo
  {journal} {Phys. Rev. A}\ }\textbf {\bibinfo {volume} {88}},\ \bibinfo
  {pages} {042105} (\bibinfo {year} {2013})}\BibitemShut {NoStop}%
\bibitem [{\citenamefont {Jensen}\ \emph {et~al.}(2007)\citenamefont {Jensen},
  \citenamefont {Kj{\ae}rgaard}, \citenamefont {Th{\o}gersen},\ and\
  \citenamefont {Fedorov}}]{BECNON}%
  \BibitemOpen
  \bibfield  {author} {\bibinfo {author} {\bibfnamefont {A.S.}\ \bibnamefont
  {Jensen}}, \bibinfo {author} {\bibfnamefont {T.}~\bibnamefont
  {Kj{\ae}rgaard}}, \bibinfo {author} {\bibfnamefont {M.}~\bibnamefont
  {Th{\o}gersen}}, \ and\ \bibinfo {author} {\bibfnamefont {D.V.}\ \bibnamefont
  {Fedorov}},\ }\bibfield  {title} {\enquote {\bibinfo {title} {Eigenvalues of
  the one-body density matrix for correlated condensates},}\ }\href
  {http://www.sciencedirect.com/science/article/pii/S0375947407004095}
  {\bibfield  {journal} {\bibinfo  {journal} {Nucl. Phys. A}\ }\textbf
  {\bibinfo {volume} {790}},\ \bibinfo {pages} {723c -- 727c} (\bibinfo {year}
  {2007})}\BibitemShut {NoStop}%
\bibitem [{\citenamefont {Armstrong}\ \emph {et~al.}(2011)\citenamefont
  {Armstrong}, \citenamefont {Zinner}, \citenamefont {Fedorov},\ and\
  \citenamefont {Jensen}}]{ZinnerNbody}%
  \BibitemOpen
  \bibfield  {author} {\bibinfo {author} {\bibfnamefont {J.R.}\ \bibnamefont
  {Armstrong}}, \bibinfo {author} {\bibfnamefont {N.T.}\ \bibnamefont
  {Zinner}}, \bibinfo {author} {\bibfnamefont {D.V.}\ \bibnamefont {Fedorov}},
  \ and\ \bibinfo {author} {\bibfnamefont {A.S.}\ \bibnamefont {Jensen}},\
  }\bibfield  {title} {\enquote {\bibinfo {title} {Analytic harmonic approach
  to the n-body problem},}\ }\href
  {http://stacks.iop.org/0953-4075/44/i=5/a=055303} {\bibfield  {journal}
  {\bibinfo  {journal} {J. Phys. B}\ }\textbf {\bibinfo {volume} {44}},\
  \bibinfo {pages} {055303} (\bibinfo {year} {2011})}\BibitemShut {NoStop}%
\bibitem [{\citenamefont {Gajda}(2006)}]{HarmBEC}%
  \BibitemOpen
  \bibfield  {author} {\bibinfo {author} {\bibfnamefont {M.}~\bibnamefont
  {Gajda}},\ }\bibfield  {title} {\enquote {\bibinfo {title} {Criterion for
  {B}ose-{E}instein condensation in a harmonic trap in the case with attractive
  interactions},}\ }\href {\doibase 10.1103/PhysRevA.73.023603} {\bibfield
  {journal} {\bibinfo  {journal} {Phys. Rev. A}\ }\textbf {\bibinfo {volume}
  {73}},\ \bibinfo {pages} {023603} (\bibinfo {year} {2006})}\BibitemShut
  {NoStop}%
\bibitem [{\citenamefont {Laughlin}(1983)}]{Laughlin1983}%
  \BibitemOpen
  \bibfield  {author} {\bibinfo {author} {\bibfnamefont {R.~B.}\ \bibnamefont
  {Laughlin}},\ }\bibfield  {title} {\enquote {\bibinfo {title} {Anomalous
  quantum {H}all effect: An incompressible quantum fluid with fractionally
  charged excitations},}\ }\href {\doibase 10.1103/PhysRevLett.50.1395}
  {\bibfield  {journal} {\bibinfo  {journal} {Phys. Rev. Lett.}\ }\textbf
  {\bibinfo {volume} {50}},\ \bibinfo {pages} {1395--1398} (\bibinfo {year}
  {1983})}\BibitemShut {NoStop}%
\bibitem [{\citenamefont {Pipek}\ and\ \citenamefont {Nagy}(2009)}]{Nagydual1}%
  \BibitemOpen
  \bibfield  {author} {\bibinfo {author} {\bibfnamefont {J.}~\bibnamefont
  {Pipek}}\ and\ \bibinfo {author} {\bibfnamefont {I.}~\bibnamefont {Nagy}},\
  }\bibfield  {title} {\enquote {\bibinfo {title} {Measures of spatial
  entanglement in a two-electron model atom},}\ }\href {\doibase
  10.1103/PhysRevA.79.052501} {\bibfield  {journal} {\bibinfo  {journal} {Phys.
  Rev. A}\ }\textbf {\bibinfo {volume} {79}},\ \bibinfo {pages} {052501}
  (\bibinfo {year} {2009})}\BibitemShut {NoStop}%
\bibitem [{\citenamefont {Schilling}\ and\ \citenamefont
  {Schilling}(2014)}]{duality}%
  \BibitemOpen
  \bibfield  {author} {\bibinfo {author} {\bibfnamefont {C.}~\bibnamefont
  {Schilling}}\ and\ \bibinfo {author} {\bibfnamefont {R.}~\bibnamefont
  {Schilling}},\ }\bibfield  {title} {\enquote {\bibinfo {title} {Duality of
  reduced density matrices and their eigenvalues},}\ }\href
  {http://stacks.iop.org/1751-8121/47/i=41/a=415305} {\bibfield  {journal}
  {\bibinfo  {journal} {J. Phys. A}\ }\textbf {\bibinfo {volume} {47}},\
  \bibinfo {pages} {415305} (\bibinfo {year} {2014})}\BibitemShut {NoStop}%
\bibitem [{\citenamefont {Ko{\'s}cik}(2015)}]{koscik2015neumann}%
  \BibitemOpen
  \bibfield  {author} {\bibinfo {author} {\bibfnamefont {P.}~\bibnamefont
  {Ko{\'s}cik}},\ }\bibfield  {title} {\enquote {\bibinfo {title} {The von
  {N}eumann entanglement entropy for {W}igner-crystal states in one dimensional
  n-particle systems},}\ }\href
  {http://www.sciencedirect.com/science/article/pii/S0375960114012171}
  {\bibfield  {journal} {\bibinfo  {journal} {Phys. Lett. A}\ }\textbf
  {\bibinfo {volume} {379}},\ \bibinfo {pages} {293--298} (\bibinfo {year}
  {2015})}\BibitemShut {NoStop}%
\bibitem [{\citenamefont {Calogero}(1969)}]{CS1}%
  \BibitemOpen
  \bibfield  {author} {\bibinfo {author} {\bibfnamefont {F.}~\bibnamefont
  {Calogero}},\ }\bibfield  {title} {\enquote {\bibinfo {title} {Ground state
  of a one-dimensional n-body system},}\ }\href
  {http://scitation.aip.org/content/aip/journal/jmp/10/12/10.1063/1.1664821}
  {\bibfield  {journal} {\bibinfo  {journal} {J. Math. Phys.}\ }\textbf
  {\bibinfo {volume} {10}},\ \bibinfo {pages} {2197--2200} (\bibinfo {year}
  {1969})}\BibitemShut {NoStop}%
\bibitem [{\citenamefont {Calogero}(1971)}]{CS2}%
  \BibitemOpen
  \bibfield  {author} {\bibinfo {author} {\bibfnamefont {F.}~\bibnamefont
  {Calogero}},\ }\bibfield  {title} {\enquote {\bibinfo {title} {Solution of
  the one-dimensional n-body problems with quadratic and/or inversely quadratic
  pair potentials},}\ }\href
  {http://scitation.aip.org/content/aip/journal/jmp/12/3/10.1063/1.1665604}
  {\bibfield  {journal} {\bibinfo  {journal} {J. Math. Phys.}\ }\textbf
  {\bibinfo {volume} {12}},\ \bibinfo {pages} {419--436} (\bibinfo {year}
  {1971})}\BibitemShut {NoStop}%
\bibitem [{\citenamefont {Sutherland}(1971{\natexlab{a}})}]{CS3}%
  \BibitemOpen
  \bibfield  {author} {\bibinfo {author} {\bibfnamefont {B.}~\bibnamefont
  {Sutherland}},\ }\bibfield  {title} {\enquote {\bibinfo {title} {Quantum
  many-body problem in one dimension: Ground state},}\ }\href
  {http://scitation.aip.org/content/aip/journal/jmp/12/2/10.1063/1.1665584}
  {\bibfield  {journal} {\bibinfo  {journal} {J. Math. Phys.}\ }\textbf
  {\bibinfo {volume} {12}},\ \bibinfo {pages} {246--250} (\bibinfo {year}
  {1971}{\natexlab{a}})}\BibitemShut {NoStop}%
\bibitem [{\citenamefont {Sutherland}(1971{\natexlab{b}})}]{CS4}%
  \BibitemOpen
  \bibfield  {author} {\bibinfo {author} {\bibfnamefont {B.}~\bibnamefont
  {Sutherland}},\ }\bibfield  {title} {\enquote {\bibinfo {title} {Exact
  results for a quantum many-body problem in one dimension},}\ }\href {\doibase
  10.1103/PhysRevA.4.2019} {\bibfield  {journal} {\bibinfo  {journal} {Phys.
  Rev. A}\ }\textbf {\bibinfo {volume} {4}},\ \bibinfo {pages} {2019} (\bibinfo
  {year} {1971}{\natexlab{b}})}\BibitemShut {NoStop}%
\bibitem [{\citenamefont {Sutherland}(1972)}]{CS5}%
  \BibitemOpen
  \bibfield  {author} {\bibinfo {author} {\bibfnamefont {B.}~\bibnamefont
  {Sutherland}},\ }\bibfield  {title} {\enquote {\bibinfo {title} {Exact
  results for a quantum many-body problem in one dimension. ii},}\ }\href
  {\doibase 10.1103/PhysRevA.5.1372} {\bibfield  {journal} {\bibinfo  {journal}
  {Phys. Rev. A}\ }\textbf {\bibinfo {volume} {5}},\ \bibinfo {pages} {1372}
  (\bibinfo {year} {1972})}\BibitemShut {NoStop}%
\bibitem [{\citenamefont {Weidem{\"u}ller}\ and\ \citenamefont
  {Zimmermann}(2011)}]{ultracoldBook1}%
  \BibitemOpen
  \bibfield  {author} {\bibinfo {author} {\bibfnamefont {M.}~\bibnamefont
  {Weidem{\"u}ller}}\ and\ \bibinfo {author} {\bibfnamefont {C.}~\bibnamefont
  {Zimmermann}},\ }\href@noop {} {\emph {\bibinfo {title} {Interactions in
  ultracold gases: from atoms to molecules}}}\ (\bibinfo  {publisher} {John
  Wiley \& Sons},\ \bibinfo {year} {2011})\BibitemShut {NoStop}%
\bibitem [{\citenamefont {Bloch}\ \emph {et~al.}(2008)\citenamefont {Bloch},
  \citenamefont {Dalibard},\ and\ \citenamefont {Zwerger}}]{BlochReview}%
  \BibitemOpen
  \bibfield  {author} {\bibinfo {author} {\bibfnamefont {I.}~\bibnamefont
  {Bloch}}, \bibinfo {author} {\bibfnamefont {J.}~\bibnamefont {Dalibard}}, \
  and\ \bibinfo {author} {\bibfnamefont {W.}~\bibnamefont {Zwerger}},\
  }\bibfield  {title} {\enquote {\bibinfo {title} {Many-body physics with
  ultracold gases},}\ }\href {\doibase 10.1103/RevModPhys.80.885} {\bibfield
  {journal} {\bibinfo  {journal} {Rev. Mod. Phys.}\ }\textbf {\bibinfo {volume}
  {80}},\ \bibinfo {pages} {885--964} (\bibinfo {year} {2008})}\BibitemShut
  {NoStop}%
\bibitem [{\citenamefont {Chin}\ \emph {et~al.}(2010)\citenamefont {Chin},
  \citenamefont {Grimm}, \citenamefont {Julienne},\ and\ \citenamefont
  {Tiesinga}}]{Feshbach}%
  \BibitemOpen
  \bibfield  {author} {\bibinfo {author} {\bibfnamefont {C.}~\bibnamefont
  {Chin}}, \bibinfo {author} {\bibfnamefont {R.}~\bibnamefont {Grimm}},
  \bibinfo {author} {\bibfnamefont {P.}~\bibnamefont {Julienne}}, \ and\
  \bibinfo {author} {\bibfnamefont {E.}~\bibnamefont {Tiesinga}},\ }\bibfield
  {title} {\enquote {\bibinfo {title} {Feshbach resonances in ultracold
  gases},}\ }\href {\doibase 10.1103/RevModPhys.82.1225} {\bibfield  {journal}
  {\bibinfo  {journal} {Rev. Mod. Phys.}\ }\textbf {\bibinfo {volume} {82}},\
  \bibinfo {pages} {1225--1286} (\bibinfo {year} {2010})}\BibitemShut {NoStop}%
\bibitem [{\citenamefont {Truscott}\ \emph {et~al.}(2001)\citenamefont
  {Truscott}, \citenamefont {Strecker}, \citenamefont {McAlexander},
  \citenamefont {Partridge},\ and\ \citenamefont {Hulet}}]{PauliPressure}%
  \BibitemOpen
  \bibfield  {author} {\bibinfo {author} {\bibfnamefont {A.G.}\ \bibnamefont
  {Truscott}}, \bibinfo {author} {\bibfnamefont {K.E.}\ \bibnamefont
  {Strecker}}, \bibinfo {author} {\bibfnamefont {W.I.}\ \bibnamefont
  {McAlexander}}, \bibinfo {author} {\bibfnamefont {G.B.}\ \bibnamefont
  {Partridge}}, \ and\ \bibinfo {author} {\bibfnamefont {R.G.}\ \bibnamefont
  {Hulet}},\ }\bibfield  {title} {\enquote {\bibinfo {title} {Observation of
  fermi pressure in a gas of trapped atoms},}\ }\href {\doibase
  10.1126/science.1059318} {\bibfield  {journal} {\bibinfo  {journal}
  {Science}\ }\textbf {\bibinfo {volume} {291}},\ \bibinfo {pages} {2570--2572}
  (\bibinfo {year} {2001})}\BibitemShut {NoStop}%
\bibitem [{\citenamefont {Z\"urn}\ \emph {et~al.}(2012)\citenamefont {Z\"urn},
  \citenamefont {Serwane}, \citenamefont {Lompe}, \citenamefont {Wenz},
  \citenamefont {Ries}, \citenamefont {Bohn},\ and\ \citenamefont
  {Jochim}}]{Jochim2species}%
  \BibitemOpen
  \bibfield  {author} {\bibinfo {author} {\bibfnamefont {G.}~\bibnamefont
  {Z\"urn}}, \bibinfo {author} {\bibfnamefont {F.}~\bibnamefont {Serwane}},
  \bibinfo {author} {\bibfnamefont {T.}~\bibnamefont {Lompe}}, \bibinfo
  {author} {\bibfnamefont {A.~N.}\ \bibnamefont {Wenz}}, \bibinfo {author}
  {\bibfnamefont {M.~G.}\ \bibnamefont {Ries}}, \bibinfo {author}
  {\bibfnamefont {J.~E.}\ \bibnamefont {Bohn}}, \ and\ \bibinfo {author}
  {\bibfnamefont {S.}~\bibnamefont {Jochim}},\ }\bibfield  {title} {\enquote
  {\bibinfo {title} {Fermionization of two distinguishable fermions},}\ }\href
  {\doibase 10.1103/PhysRevLett.108.075303} {\bibfield  {journal} {\bibinfo
  {journal} {Phys. Rev. Lett.}\ }\textbf {\bibinfo {volume} {108}},\ \bibinfo
  {pages} {075303} (\bibinfo {year} {2012})}\BibitemShut {NoStop}%
\bibitem [{\citenamefont {Lieb}(1983)}]{LiebDFT}%
  \BibitemOpen
  \bibfield  {author} {\bibinfo {author} {\bibfnamefont {E.H.}\ \bibnamefont
  {Lieb}},\ }\bibfield  {title} {\enquote {\bibinfo {title} {Density
  functionals for {C}oulomb systems},}\ }\href {\doibase 10.1002/qua.560240302}
  {\bibfield  {journal} {\bibinfo  {journal} {Int. J. Quant. Chem.}\ }\textbf
  {\bibinfo {volume} {24}},\ \bibinfo {pages} {243--277} (\bibinfo {year}
  {1983})}\BibitemShut {NoStop}%
\bibitem [{\citenamefont {Parr}\ and\ \citenamefont {Yang}(1994)}]{DFTbook1}%
  \BibitemOpen
  \bibfield  {author} {\bibinfo {author} {\bibfnamefont {R.G.}\ \bibnamefont
  {Parr}}\ and\ \bibinfo {author} {\bibfnamefont {W.}~\bibnamefont {Yang}},\
  }\href
  {http://www.oup.com/us/catalog/general/subject/Chemistry/TheoreticalChemistry/?view=usa\&\#38;ci=9780195092769}
  {\emph {\bibinfo {title} {{Density-Functional Theory of Atoms and
  Molecules}}}}\ (\bibinfo  {publisher} {Oxford University Press},\ \bibinfo
  {year} {1994})\BibitemShut {NoStop}%
\bibitem [{\citenamefont {Gradshteyn}\ and\ \citenamefont
  {Ryzhik}(2007)}]{gradshteyn2007}%
  \BibitemOpen
  \bibfield  {author} {\bibinfo {author} {\bibfnamefont {I.~S.}\ \bibnamefont
  {Gradshteyn}}\ and\ \bibinfo {author} {\bibfnamefont {I.~M.}\ \bibnamefont
  {Ryzhik}},\ }\href
  {http://www.lepp.cornell.edu/~ib38/tmp/reading/Table_of_Integrals_Series_and_Products_Tablicy_Integralov_Summ_Rjadov_I_Proizvedennij_Engl._2.pdf}
  {\emph {\bibinfo {title} {Table of integrals, series, and products}}},\
  \bibinfo {edition} {seventh}\ ed.\ (\bibinfo  {publisher} {Elsevier/Academic
  Press, Amsterdam},\ \bibinfo {year} {2007})\BibitemShut {NoStop}%
\end{thebibliography}%

\onecolumngrid
\begin{appendix}
\setcounter{secnumdepth}{1}
\section{Calculation of $F(x,y)$ and $V_N(\tilde{p})$ for $t \rightarrow 0$}\label{app:FandV}
\renewcommand{\theequation}{\thesection\arabic{equation}}
\setcounter{equation}{0}

In this appendix we calculate the leading order in $t$ of the polynomial $F(x,y)$ and of the effective potential $V_{N}(\tilde{p})$. $F(x,y)$ is given by
\begin{equation} \label{eqA1}
F(x,y) =\int\limits^\infty _{- \infty} d u \; e ^{- u^2 } \sum\limits_{k=0}^{N-1} \frac{1}{2^k k!} H_k (p u + q (x,y)) H_k (p u + q (y, x))\,,
\end{equation}
where $H_k (x)$ are the Hermite polynomials and
\begin{equation} \label{eqA2}
p=[B / (A-(N-1)B)]^{1/2}\,,\quad  q(x,y)=\sqrt{2A} \Big[x-\frac{1}{2} p^2 (x +y) \Big] \, .
\end{equation}
From Eq.~(\ref{eqA2}) we obtain with Eq.~(\ref{eq3}) and $t=l_+/l_-$
\begin{equation} \label{eqA4}
p = 1 + O(t^2)\,,\quad q(x,y) =\frac{1}{2} \Big(1 + O(t^2) \Big) \frac{x-y}{t} + O(t) x \, .
\end{equation}
The following identity
\begin{equation} \label{eqA6}
H_k(x+y) = \sum\limits_{m=0}^k \Big({k \atop m}\Big) H_m (x) (2y)^{k-m}
\end{equation}
is useful. It can be proved by Taylor-expanding its left hand side with respect to $y$
and using for the $n$-th derivative $H_{k}^{(n)}(x)=2^{n}\frac{k!}{(k-n)!}H_{k-n}(x)$ which is easily proved by induction taking the initial condition $H_{k}^{(1)}(x)=2kH_{k-1}(x)$ \cite{gradshteyn2007} into account.
Substituting the leading order term of $p$ and $q(x,y)$ into Eq.~(\ref{eqA1})
and making use of the identity (\ref{eqA6})
one gets
\begin{eqnarray} \label{eqA7}
 F(x,y) &\cong &\sum\limits_{k=0}^{N-1} \frac{1}{2^k k!} \sum\limits _{m=0}^k \;\sum\limits_{m'=0}^k \Big( {k \atop m}\Big) \Big({k \atop m'} \Big)  \Big(\frac{x-y}{t} \Big)^{k-m} \Big(\frac{y-x}{t}\Big)^{k-m'} \nonumber \\
 && \cdot \int\limits^\infty_{- \infty} du\; e^{-u^2} H_m (u) H_{m'} (u) \nonumber\\
  &=& \sqrt{\pi} \sum\limits_{k=0}^{N-1} \; \frac{1}{2^k k!} \; \sum\limits_{m=0}^k \Big({k \atop m} \Big)^2 2 ^m m! \Big(-1 \Big) ^{k-m} \Big(\frac{x-y}{t}\Big)^{2(k-m)} \nonumber\\
  &=& \sqrt{\pi} \sum\limits_{m=0}^{N-1} \frac{(-1)^m}{2^m m!} \Big[\sum\limits_{k=m}^{N-1} \Big({k \atop m}\Big)  \Big] \Big(\frac{x-y}{t} \Big) ^{2m} \nonumber\\
&=& \sqrt{\pi} \sum\limits_{m=0}^{N-1} \frac{(-1)^m}{2^m m!} \Big({N \atop m+1}\Big)  \Big(\frac{x-y}{t} \Big) ^{2m} \nonumber\\
&\equiv& \tilde{F}(\tilde{z})\,,
\end{eqnarray}
where $\tilde{z}=(x-y)/t$. We have used  $\int\limits_{- \infty}^\infty du \; e^{- u^2} H_m(u) H_{m'} (u) =\sqrt{\pi} 2 ^m m! \delta_{mm'}$ \cite{gradshteyn2007}
and $\sum\limits_{k=m}^{N-1} \Big({k \atop m}\Big)= \Big({N \atop m+1}\Big)$ \cite{gradshteyn2007}. Note that $\tilde{F}(\tilde{z})$ is an even function in $\tilde{z}$.
It is straightforward to calculate  $\langle \tilde{z}^{2n}\tilde{F}(\tilde{z})\rangle =\int\limits^\infty_{- \infty} d \tilde{z} \tilde{z}^{2n}\tilde{F}(\tilde{z}) \exp{(- \frac{N-1}{4N} \tilde{z}^2)}$. One obtains
\begin{equation} \label{eqA8}
 \langle \tilde{z}^{2n}\tilde{F}(\tilde{z})\rangle = \pi \sqrt{4N/(N-1)}  \sum\limits_{m=0}^{N-1}(-1)^{m}\frac{(2(n+m)-1)!!}{2^{m}m!}\Big({N \atop m+1}\Big) \Big(\frac{2N}{N-1}\Big)^{m+n} \, .
\end{equation}
Making use of Eq.~(\ref{eqA7}) one can also calculate the effective potential (Eq.~(\ref{eq13})) which can be represented as $V_N(\tilde{p})=\langle \tilde{F}(\tilde{z})\cos(\sqrt{t}\tilde{p}\tilde{z})\exp(-\frac{N-1}{4N} \tilde{z}^2)\rangle$. One obtains with $\mathcal{N} \simeq \sqrt{N}/\pi$(cf.~Eq.~(\ref{eqB3}))
\begin{equation} \label{eqA9}
V_N(\tilde{p})=-t\Big[\sum\limits_{n=0}^{N-1}(-1)^{n}v_{N,n} \Big(\frac{2tN}{N-1}\tilde{p}^2\Big)^{n} \Big] \exp(-\frac{t N}{N-1}\tilde{p}^2)
\end{equation}
with coefficients
\begin{equation} \label{eqA10}
v_{N,n}=\frac{2N}{\sqrt{N-1}}\sum\limits_{m=n}^{N-1}(-1)^{m}\frac{(2(m-n)-1)!!}{2^{m}m!} \Big({N \atop m+1}\Big)\Big({2m \atop 2n}\Big) \Big(\frac{2N}{N-1}\Big)^m \,.
\end{equation}

\section{Mapping of the $1$-RDM for $t \to 0$}\label{app:1rdm}
This appendix presents details of the reduction of the eigenvalue equation
(\ref{eq10}) for the $1$-RDM in the strong coupling limit $t \to 0$ to an eigenvalue equation for a particle in an effective potential $V_N(\tilde{p})$.
Using Eqs.~(\ref{eq3}) and~(\ref{eq8}), one obtains for the exponent on the right hand side of Eq.~(\ref{eq4})
\begin{equation} \label{eqB1}
 a(x^2 + y^2) - b xy=\frac{N-1}{4N} \Big(1 + O (t^2)\Big) \Big(\frac{x-y}{t}\Big)  +N(1 + O (t^2)) x y \quad.
\end{equation}
From Eq~(\ref{eqA7}) we have
\begin{equation} \label{eqB2}
F(x,y) \cong \sqrt{\pi} \sum\limits_{m=0}^{N-1} \frac{(-1)^m}{2^m m!} \Big({N \atop m+1}\Big)  \Big(\frac{x-y}{t} \Big) ^{2m}
\equiv \tilde{F} \Big(\frac{x-y}{t} \Big) \quad.
 \end{equation}
Taking into account that Eqs.~(\ref{eqB1}) and (\ref{eqB2}) imply   $a (x^2+y^2) -b xy]_{x=y} \cong N x^2$ and $F(x,x) \cong \sqrt{\pi} N$, respectively,  the normalization condition for $\rho_1 (x;y)$ gives the normalization constant in leading order in $t$
\begin{equation} \label{eqB3}
 \mathcal{N} \cong \frac{\sqrt{N}}{\pi} \quad .
 \end{equation}
The results~(\ref{eqB1}) and ~(\ref{eqB2}) show that $\rho_1 (x; y)$ in leading order only depends on $(x-y)$, i.e.~is translationally invariant. This can easily be understood since for $l_+$  fixed, $t \rightarrow 0$ implies that the length scale $l_-$ associated with the harmonic trap converges to infinity, i.e.~its eigenfrequency $\omega$ goes to zero. Consequently, the external potential becomes flat for $t \rightarrow 0$, and the hamiltonian (\ref{eq1}) becomes invariant under translations.
To proceed, we introduce scaled variables,
\begin{equation} \label{eqB4}
\tilde{x} = x/\sqrt{t} \quad , \quad \tilde{y}=y/ \sqrt{t} \quad, \quad \tilde{z}=(x-y)/t= (\tilde{x}-\tilde{y})/\sqrt{t}\,.
\end{equation}
Then we get from Eqs.~(\ref{eq4}), (\ref{eqB1}) - (\ref{eqB4})
\begin{eqnarray} \label{eqB5}
\rho_1(x;y) \cong \mathcal{N}\exp{(-N t \tilde{x}^2)}\exp{(Nt^{\frac{3}{2}}\tilde{x}\tilde{z})} \tilde{F} (\tilde{z}) \exp \Big[-\frac{N-1}{4N} \tilde{z}^2 \Big] \,.
\end{eqnarray}
Going beyond the leading order term ~(\ref{eqB2}) of $F(x,y)$ there also appear additional terms $((x-y)/t)^m = \tilde{z}^m$ for $m=0,1, \ldots, N-1$ with coefficients of order $t^2$ and terms $\tilde{x} (\tilde{x}-\sqrt{t} \tilde{z})$ with a coefficient of order $t^3$. Therefore the right hand side of Eq.~(\ref{eqB5}) with $F(x,y)$ given by Eq.~(\ref{eqB2}) represents the leading order terms of the expansion of $\rho_1 (x,y)$ for $t \rightarrow 0$.

Next, we introduce scaled natural orbitals
\begin{equation} \label{eqB6}
\tilde{\chi}_k (\tilde{x}) = \chi_k (\sqrt{t} \tilde{x}) \quad.
\end{equation}
By making use of Eqs.~(\ref{eqB3})-(\ref{eqB6}) and choosing $\tilde{z}$ as an integration variable the eigenvalue equation~(\ref{eq10}) becomes with
$(N-1)/(4N))=\alpha$
\begin{equation} \label{eqB7}
t \ \mathcal{N}\int\limits_{- \infty}^\infty d \tilde{z} \exp{(-N t \tilde{x}^2)}\exp{(Nt^{\frac{3}{2}}\tilde{x}\tilde{z})}\tilde{F} (\tilde{z}) e^{-\alpha \tilde{z}^2} \tilde{\chi}_k (\tilde{x}-\sqrt{t} \tilde{z}) =\lambda_k \tilde{\chi}_k (\tilde{x}) \,.
\end{equation}
The second exponential in Eq.~(\ref{eqB7}) with the exponent involving $\tilde{x}\tilde{z}$ can be eliminated by using
\begin{equation} \label{eqB8}
\tilde{\chi}_k (\tilde{x})=\exp{(\frac{1}{2}Nt\tilde{x^2})}\zeta_k(\tilde{x}) \ .
\end{equation}
This leads to
\begin{equation} \label{eqB9}
t \ \mathcal{N} \exp{(-N t \tilde{x}^2)}\int\limits_{- \infty}^\infty d \tilde{z}\tilde{F} (\tilde{z}) e^{- \alpha \tilde{z}^2} \zeta_k (\tilde{x}
-\sqrt{t} \tilde{z})= \lambda_k \zeta_k (\tilde{x}) \,,
\end{equation}
where $\exp{(\frac{1}{2}Nt\tilde{z}^2)}$ (which occurs in the integrand) can be neglected with respect to $\exp{(- \alpha \tilde{z}^2)}$, since $\alpha=O(t^{0})$.

Next we use
\begin{equation} \label{eqB10}
\zeta_k (\tilde{x}-\sqrt{t} \tilde{z})= \exp{\Big(-\sqrt{t} \tilde{z}\frac{\partial}{\partial \tilde{x}}\Big) }\zeta_k (\tilde{x}) \ .
\end{equation}
Substituting Eq.~(\ref{eqB10}) into  Eq.~(\ref{eqB9}) and taking into account that $F (\tilde{z})$ is even in $\tilde{z}$ we get with the momentum operator
$\hat{\tilde{p}}=\frac{\hbar}{i}\frac{\partial}{\partial \tilde{x}}$
\begin{equation} \label{eqB11}
t \ \mathcal{N} \exp{(-N t \tilde{x}^2)} \Big[\int\limits_{- \infty}^\infty d \tilde{z}\tilde{F} (\tilde{z}) e^{- \alpha \tilde{z}^2} \cos{(\sqrt{t} \tilde{z}\hat{\tilde{p}})} \Big] \zeta_k (\tilde{x}) =\lambda_k \zeta_k (\tilde{x}) \,.
\end{equation}
Using  momentum representation, i.e.~Fourier transforming $\zeta_k (\tilde{x})$
\begin{equation} \label{eqB12}
\zeta_k (\tilde{x}) = \frac{1}{2\pi}\int\limits_{- \infty}^\infty d \tilde{p} \ \tilde{\zeta}_k (\tilde{p})  e^{-i\tilde{x}\tilde{p}}\ \quad ,
\end{equation}
turns out to be the appropriate choice.
Then Eq.~(\ref{eqB11}) becomes
\begin{equation} \label{eqB13}
\exp{(N t \frac{\partial^2}{\partial \tilde{p}^2})} \Big[(-t  \mathcal{N})\int\limits_{- \infty}^\infty d \tilde{z}\tilde{F} (\tilde{z})  \cos{(\sqrt{t} \tilde{z}\tilde{p})} e^{- \alpha \tilde{z}^2} \Big] \tilde{\zeta}_k (\tilde{p}) =(-\lambda_k) \tilde{\zeta}_k (\tilde{p}) \, .
\end{equation}
Note, the term in the square bracket is just the effective potential $V_N(\tilde{p})$ from Eq.~(\ref{eq13}).
The final step is the expansion of $\exp{(N t \frac{\partial^2}{\partial \tilde{p}^2}})$.
This yields $\exp{(N t \frac{\partial^2}{\partial \tilde{p}^2})V_N(\tilde{p}})
=[1+N t \frac{\partial^2}{\partial \tilde{p}^2}+O(t^2)]V_N(\tilde{p}) \simeq V_N(\tilde{p})+V_N(\tilde{p}=0)N t \frac{\partial^2}{\partial \tilde{p}^2}$. Substituting this into Eq.~(\ref{eqB13}) leads immediately to the eigenvalue equation (\ref{eq11}) with the mass $\tilde{m}$ given by Eq.~(\ref{eq12}).

Finally, we justify the scaling (\ref{eqB6}) and the validity of the implicit assumption that
$\tilde{\chi}_k$ varies on a length scale $\tilde{x}=O(1)$.
From  Eq.~(\ref{eq13}) one gets $V_N(0)=O(t)$,
$V''^{(min)}_{N}=O(t^2)$
which leads to $\tilde{m}=O(t^{-2})$ and
$\tilde{\Omega}=\sqrt{V''^{(min)}_{N}/ \tilde{m}}=O(t^2)$. The length scale corresponding to the harmonic oscillator equation obtained from Eq.~(\ref{eq11}) by expanding $V_N(\tilde{p})$
up to second order in $(\tilde{p}-\tilde{p}_{min})$ is given by
$\tilde{L}=\sqrt{\frac{\hbar}{\tilde{m}\tilde{\Omega}}}$ which is $O(1)$. Accordingly, $\tilde{\zeta}_k$, and therefore $\tilde{\chi_k}$, varies on a scale of $O(1)$.

\section{Calculation of $\lambda_k$ for $k \gg k_{*}(t)$}\label{app:NONs}

Here we follow the strategy used in Ref.~\cite{CS2013NO}. We present the crucial steps, only. For details
the reader may consult Appendix C of Ref.~\cite{CS2013NO}. There it was shown that
$\rho_1 (x;y)\equiv \langle x|\hat{\rho}_1|y \rangle$ with the 1-particle reduced density operator

\begin{equation} \label{C1}
\hat{\rho}_1 =\sum\limits_{\nu=0}^{N-1} \; \sum\limits_{\mu=0}^{2 \nu} c_{\nu, \mu} \hat{x}^{2 \nu-\mu} \; e^{- \beta_N \hat{H}_{\rm eff}} \hat{x}^\mu
\end{equation}
and the effective hamiltonian $\hat{H}_{\rm eff}=\hbar \Omega_N \Big(a^{\dagger} a + \frac{1}{2}\Big)$. $a$ and $a^{\dagger}$  are boson annihilation and creation operators and  $c_{\nu, \mu}$ the  coefficients of the polynomial $F (x,y)$. The explicit form for $\Omega_N$  is given in Ref.~\cite{CS2013NO} and is not essential here. The position operator $\hat{x}$ (in units of $l_-$) can be represented as follows

\begin{equation} \label{C2}
\hat{x} = \frac{L^{(b)}}{\sqrt{2}} (a + a^{\dagger}) \cong \tilde{L}^{(b)}_N \sqrt{\frac{t}{2}} (a + a^{\dagger})
\end{equation}
with the bosonic length scale $L^{(b)}=\tilde{L}^{(b)}\sqrt{t}$ and $\tilde{L}^{(b)}=(N-1)^{-1/4}$ .

Let $|m \rangle$ be an eigenstate of $a^{\dagger}a$ with eigenvalue $m$ and $m\gg 1$. Then it follows for $\mu \ll m$

\begin{equation} \label{C3}
(a + a^{\dagger})^{\mu} |m \rangle \cong m^{\mu/2} \sum\limits_{\kappa=0}^\mu \Big({\mu \atop \kappa}\Big) |m + \mu - 2 \kappa \rangle \quad .
\end{equation}

Following Appendix C of Ref.~\cite{CS2013NO} and correcting a few typos one obtains

\begin{eqnarray} \label{C4}
\hat{\rho}_1 |m \rangle &\cong &\Big(\frac{1}{2} \tilde{L}_N^{(b)2} t \ m \Big)^{N-1} \sum\limits_{\nu=0}^{N-1} \Big(\frac{1}{2} \tilde{L}^{(b)2}_N t \ m \Big)^{-(N-1-\nu)} \sum\limits^{2 \nu}_{\mu=0} c_{\nu,\mu} \sum\limits^\mu_{\kappa=0} \Big({\mu \atop \kappa}\Big) \nonumber\\
&& \cdot \exp \Big[-\beta_N \hbar \Omega _N \Big(m + \mu-2 \kappa + \frac{1}{2} \Big)\Big] \sum\limits^{2 \nu - \mu}_{\tau=0} \Big({ 2 \nu-\mu \atop \tau}\Big) |m+2 (\nu-\kappa-\tau ) \rangle \nonumber\\
& \cong & \Big(\frac{1}{2} \tilde{L}^{(b)2} t \ m \Big)^{N-1} \; e^{-\beta_N \hbar \Omega_N \Big(m + \frac{1}{2}\Big)} \sum\limits_{r=-(N-1)}^{N-1} h_{m, m+r} |m + 2 r \rangle\,,
\end{eqnarray}
where $m\gg t^{-1}$ has been assumed, for fixed $N$. The coefficients $\{h_{m, m+r}\}$ can be expressed by $\{c_{\nu,\mu\}}$ and do not depend  explicitly on $m$. Therefore, the eigenvalue equation

$$\hat{\rho}_1 |\chi_k \rangle =\lambda_k |\chi_k \rangle$$
for $|\chi_k \rangle=\sum\limits_{m=0}^\infty \chi^{(m)}_k |m \rangle$ becomes a finite difference equation for the coefficient $\chi^{(m)}_k$

\begin{equation} \label{eqC5}
\Big(\frac{1}{2} \tilde{L}^{(b)2}_N t \ m \Big)^{N-1} \; e^{-\beta_N \hbar \Omega _N \Big(m + \frac{1}{2}\Big)} \sum\limits_{r= - (N-1)} ^{N+1}h_{m, m+r} \chi_k^{(m + r)}  =\lambda_k \chi_k^{(m)} \quad .
\end{equation}
In Ref.~\cite{CS2013NO} the eigenvalues $\lambda_k$ in Eq.~(\ref{eqC5})
were determined. Using the leading order result  $\beta_N \hbar \Omega _N \cong 2 N  (\tilde{L}_{N}^{(b)})^2\ t $ (which follows from Ref.~\cite{CS2013NO}) one obtains

\begin{equation}  \label{C6}
\lambda_k \cong \Big(\frac{1}{2} \tilde{L}^{(b)2}_N kt\Big)^{N-1} \exp \Big[-2N \tilde{L} ^{(b)2}_N \Big(k - \frac{1}{2} \Big) t\Big] \,,
\end{equation}
which is up to a constant identical to Eq.~(\ref{eq16}).
\end{appendix}

\end{document}